\documentclass[nhess, manuscript]{copernicus}

\usepackage{enumerate}
\usepackage{ulem} 

\usepackage{fancyhdr}
\begin{document}

\nolinenumbers

\title{Urban Flood Drifters (UFDs): Onset of Movement}

\Author[1]{Daniel}{Valero}
\Author[2,*]{Arnau}{Bayón}
\Author[1]{Mário J.}{Franca}

\affil[1]{Karlsruhe Institute of Technology (Germany)}
\affil[2]{Department of Hydraulic Engineering and Environment, Universitat Politècnica de València (Spain)}

\correspondence{Arnau Bayón (arbabar@upv.es)}

\runningtitle{Urban Flood Drifters (UFDs): Onset of Movement}

\runningauthor{Valero et al.}

\received{}
\pubdiscuss{}
\revised{}
\accepted{}
\published{}

\firstpage{1}

\maketitle

\begin{abstract}
Despite their catastrophic implications in flood events, the mobilization and transport of large, loose objects—termed Urban Flood Drifters (UFDs)—are often overlooked in flood management.
This oversight stems from our limited understanding of how flowing water interacts with these heterogeneous objects.
To bridge this knowledge gap, we introduce a mechanistic stability model that predicts the onset of UFD mobilization across a diverse array of loose objects, from plastics to heavy vehicles.
Built on an inventory of key physical properties of UFDs, this model is also validated against existing mobilization studies. Our model generates stability curves that delineate flow conditions leading to their mobilization. We further enhance the reliability of our model by incorporating a Monte Carlo-based probabilistic framework that accounts for uncertainties and interdependencies among the input parameters. These probabilistic stability curves enable us to estimate the likelihood of mobilization for specific categories of UFDs under certain flow conditions. When integrated with flood maps or two-dimensional (2D) hydrodynamic models, our stability curves can guide urban planning efforts to predict and mitigate the impacts of UFDs during extreme flood events.
\end{abstract}


\section{Introduction}  
\label{sec:Introduction}

\thispagestyle{empty}
\fancyhead[R]{\textit{This preprint is currently under revision}}

Floods exert significant impacts on human life, economic stability, environmental health, and cultural heritage  \citep{Hickey1995environmental, Jonkman2008lossoflife, Hammond2015urban, Arrighi2021UNESCO}. 
Between 2000 and 2019, floods affected 1.65 billion people globally, resulting in over 100,000 fatalities \citep{Browder2021, UN2023report}. Floods account for more than half of the total damage attributed to natural disasters since the 1980s  \citep{European2012damages}.
According to the \cite{WEF2022}, floods are one of the top three most frequent and severe weather events, predicted to displace over 200 million people by 2050 \citep{clement2021}. 
Particularly in the European Union, approximately two-thirds of the aggregate damage caused by natural disasters since the 1980s can be attributed to hydrometeorological events \citep{European2012damages}. Moreover, a single flood event can incur direct costs running into thousands of millions of euros \citep{European2012damages}. \\

Low- and middle-income countries are acutely susceptible  \citep{Kahn2005death}, and even within countries with long tradition in flood defenses, a disparity exists in flood protection preparedness across urban areas of varying resources  \citep{Lindersson2023NatSus}.
Global warming and land-use change further compound the situation by amplifying the frequency and intensity of floods \citep{UN2020Climate}, thereby further exposing humans to these disastrous events \citep{Winsemius2018disaster}. \\


\begin{figure}[H]
\centering
\includegraphics[width=1\textwidth]{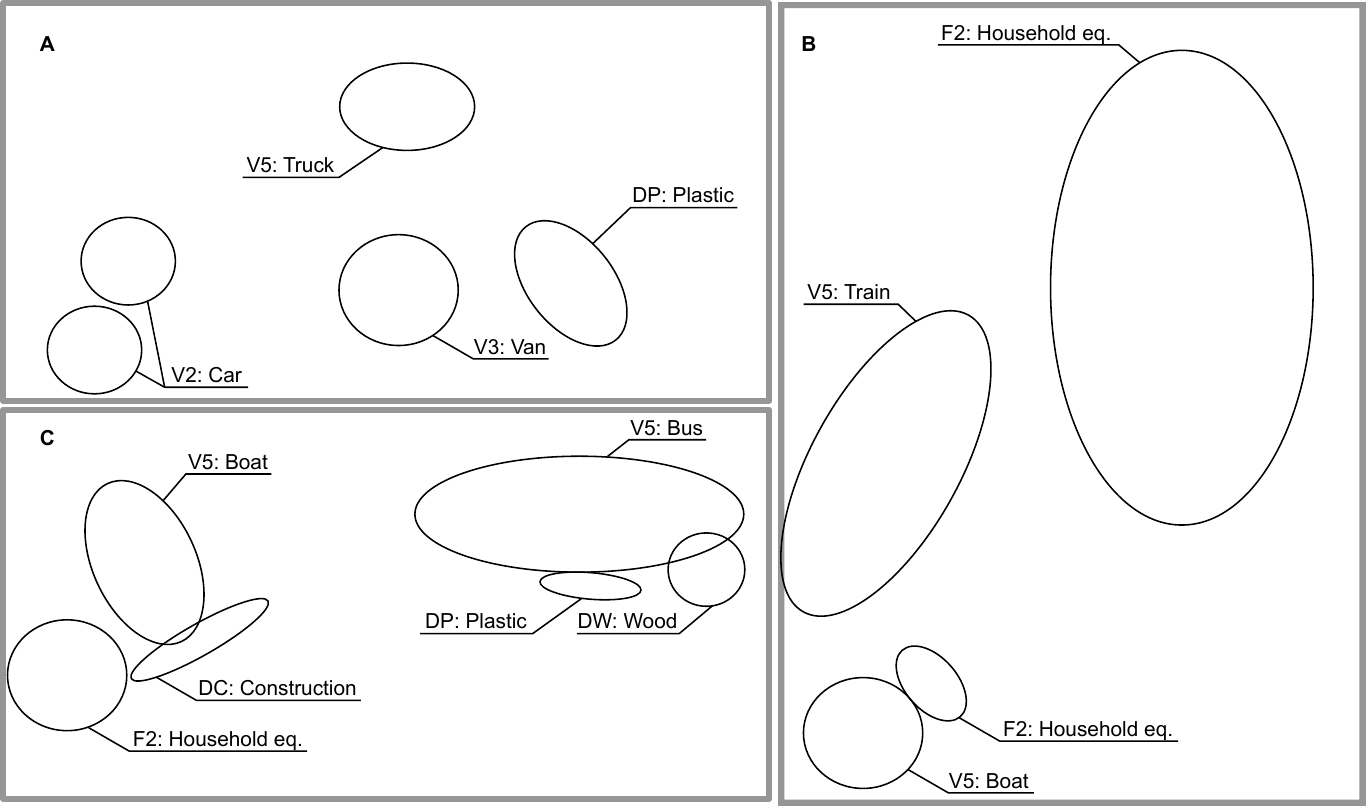}
\caption{Examples of Urban Flood Drifters (UFDs) categorized according to Table \ref{tab:Classification_UFDs}, based on \cite{Bayon2023}: \textbf{A} Sanliurfa, Turkey \citep[credit: ][]{A}, \textbf{B} Cedar Rapids, USA \citep[credit: ][]{B}, and \textbf{C} Laishui, China \citep[credit: ][]{C} }
\label{Fig:UFDs_pics}
\end{figure}

A critical yet underexamined aspect of floods is their ability to mobilize Urban Flood Drifters (UFDs, Figure \ref{Fig:UFDs_pics}), comprising loose objects commonly found in urban settings \citep{Bayon2023}. 
The resultant hazards range from structural damage to buildings  \citep{Jalayer2018debris, Zhang2018debris} to the obstruction of critical flood relieving infrastructure \citep{Kramer2015debris,davidson2015large}.
Despite growing awareness of the role of UFDs in floods \citep{Dewals2021july, Mohr2022multi, Ludwig2022multi}, this phenomenon remains largely underrepresented in urban flood research and impact evaluations.
A comprehensive hydro-mechanical framework to gauge the mobilization and transport of UFDs is still lacking, thus limiting the precise forecasts of their impact in flood scenarios.
Existing research has largely concentrated on the stability of vehicles \citep{Keller1992stability, martinez2018stability, Bocanegra2020review} and trash containers  \citep{Martinez2020containers}, often under controlled, experimental conditions. 
Such mechanistic, laboratory scale studies, have only recently been extended to incipient motion of plastic debris \citep{Waldschlaeger2019erosion,Goral2023shields,VanEmmerik2023stability}, and their transport \citep{Valero2022surfaced,Lofty2023bedload}.

In contrast, our study aims to develop a robust mechanistic model capable of addressing a broader array of UFD types, as categorized in Table \ref{tab:Classification_UFDs} following \citet{Bayon2023}.
Building on readily accessible catalog parameters, our model outlines the threshold conditions for flotation, sliding, and toppling of various UFDs, considering simple flow variables (depth $h$ and depth-averaged velocity $v$).
We evaluate the performance of the model across different car types, based on prior studies on vehicular flood instability.
Furthermore, we analyze the attribute interdependencies among different UFDs, enabling a mathematical profiling of these objects.
By adopting a Monte Carlo technique, we account for uncertainties in estimating stability limit states, thereby capturing the inherent variability associated with the onset of UFD mobilization.


\begin{table}[H]
    \centering
    \small 
    \begin{tabular}{  c  p{1.1 cm}  p{0.4cm} p{3.1 cm} p{9.65cm} }
 
        \hline
        {} & Category & ID &  Subcategory & Description\\
        \hline
        \textbf{Typified UFDs} & UFD-V & V1 & Two-wheelers & Bikes, motorbikes and e-scooters.\\
        {} & {} & V2 & Cars & Cars and other light four-wheel vehicles designed to transport of passengers.\\
        {} & {} & V3 & Vans & Vans and other heavy four-wheel vehicles designed to transport materials and stock.\\
        {} & {} & V4 & Caravans \& RVs & Vehicles designed to provide habitable space (RV: recreational vehicle).\\
        {} & {} & V5 & Large heavy vehicles & Vehicles designed to transport a large amount of people or goods (buses, trucks, trains, boats, etc.).\\
        {} & UFD-F & F1 & Urban fixtures & Facilities designed to provide a public service in streets (bins, waste containers, etc.).\\
        {} & {} & F2 & Household equipment & Facilities from private front (and back) gardens that can be carried by floods (tanks, garden sheds, etc.).\\
        \hline
        \textbf{Heterog. UFDs} & UFD-H & DC & Construction & Debris that can be dragged from construction sites or damaged buildings.\\
        {} & {} & DM & Metal & Metal debris, predominantly of constructive origin (sheets, pipes, etc.).\\
        {} & {} & DP & Plastic & Plastics and textile objects of small dimensions and irregular shape.\\
        {} & {} & DW & Wood & Natural wood (trunks, branches, etc.) and processed wood.\\
        {} & {} & DO & Others & Other drifters of uncertain origin (food, tableware, leaves, sediment, etc.). \\
        \hline
        
    \end{tabular}
    \caption{Classification and subgrouping of UFDs based on the analysis of \citet{Bayon2023}.}
    \label{tab:Classification_UFDs}
\end{table}


\section{Urban Flood Drifters (UFDs) and their key characteristics}
\label{sec:UFD_def}

Drawing upon historical data from significant flood events over the past 35 years, \cite{Bayon2023} highlighted both the diversity (Table \ref{tab:Classification_UFDs}) and prevalence of UFDs in urban environments following catastrophic flooding. This analysis distinguished two main categories of UFDs: 1) typified UFDs, which correspond to standardized items found in catalogues and are manufactured with specific dimensions and weights, and 2) a heterogeneous mixture comprising mainly of flood-damaged debris/drifters of various shapes and weights. Within the typified UFD category, two functional subcategories are discernible, namely vehicles and furniture.

\cite{Bayon2023}'s statistical evaluation reveals a considerable volume of heterogeneous debris and drifters --including plastics, construction debris, wood--, furniture (both public and private), and even heavy vehicles. A review of this study and other investigations on urban flooding \citep{Mignot2019review, Smith2019developing_countries, ODonnell2020urban, Zevenbergen2020flood, Wing2022inequitable, Bates2023climate, Sanders2023large} indicates that the stability characteristics of the majority of UFDs outlined in Table \ref{tab:Classification_UFDs} remain largely unassessed.

\begin{figure}[H]
\includegraphics[width=18cm]{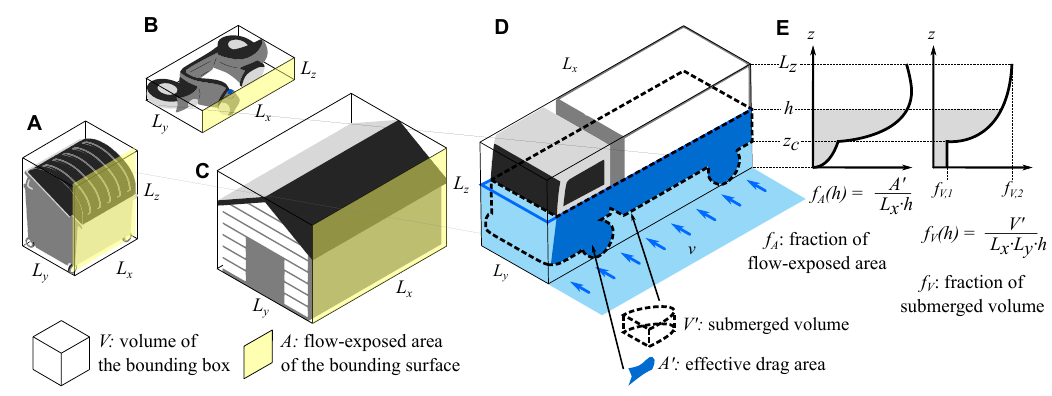}
\caption{Definition of the bounding box, with dimensions $L_x$, $L_y$ and $L_z$, over an UFD of types (\textbf{A}) UFD-F1 (urban fixture), (\textbf{B}) UFD-V1 (two-wheeler),  (\textbf{C}) UFD-F2 (household equipment) and (\textbf{D}) UFD-V5 (large heavy vehicle).
Definition of the bounding box's total volume $V$ ($=L_x L_y L_z$) and submerged volume $V'$ ($=f_V L_x L_y h$), its bounding box flood-prone area $A$ ($= L_x h$) and submerged area exposed to the flow $A'$ ($= f_A L_x h$), and its area fraction ($f_A$) and volume fraction ($f_V$).}
\label{Fig:BoundingBox_fA_fV}
\end{figure}

Earlier research frequently employed experimental setups, often involving scale models of cars, to analyze stability curves. These curves are derived from varying water depths ($h$) and depth-averaged flow velocities ($v$) under controlled hydraulic laboratory conditions \citep[see, for example,][for 1:1 scale models]{Kramer2016safety, Smith2019full}. Furthermore, theoretical models grounded in first principles have been developed and calibrated against experimental data to predict unstable conditions for specific car models \citep{Xia2014criterion, Shah2021criterion} and trash bins \citep{Martinez2020containers}. Following \cite{Martinez2020containers} and \cite{Smith2019full}, among others, we identify several key characteristics of UFDs that influence their dynamics during flood events:
\begin{enumerate}
    \item The bounding box dimensions ($L_x$, $L_y$, $L_z$, Fig. \ref{Fig:BoundingBox_fA_fV}A-D), which tightly contain a UFD, with $L_x$ denoting the longest horizontal length, $L_y$ the shortest horizontal length, and $L_z$ the vertical height (floor-normal).
    
    \item The total mass ($M$) of the UFD, which can be reciprocally defined using the UFD's bulk density ($\rho_b$):
    \begin{equation}
        \rho_b = \frac{M}{V}
        \label{eq:rho_b}
    \end{equation}
    where $V=L_x L_y L_z$ is the enclosed volume of the UFD bounding box.

    \item The submerged volume ($V'$), which is defined as the volume contributing to buoyancy when submerged up to a water depth $h$. This is inherently smaller than $V$ and is formulated as a fraction thereof (see Fig. \ref{Fig:BoundingBox_fA_fV}E):
    \begin{equation}
        f_V (h) = \frac{V'}{L_x L_y h}
        \label{eq:f_V}
    \end{equation}
    This ratio closely resembles porosity. A value of $f_V \approx 1$ suggests a fully water-tight cubic volume occupying the entirety of the volume defined by $L_x L_y h$.
        
    \item The flow-exposed area of the bounding box, given by $A = L_x h$. For two-wheelers like scooters and bikes, which are assumed to lie on their sides (Fig. \ref{Fig:BoundingBox_fA_fV}B), this assumption is based on the expectation that during the initial stages of flooding, two-wheelers may often topple, settling into a more stable, horizontal position.  
    
    \item The UFD effective drag area, which is the actual side area contributing to drag and is only a fraction ($f_A$) of the flow-exposed bounding box area (Fig. \ref{Fig:BoundingBox_fA_fV}E):
    \begin{equation}
        f_A (h) = \frac{A'}{L_x h}
        \label{eq:f_A}
    \end{equation}
    Here, an $f_A$ value close to 1 implies that the UFD effectively obstructs most of the water flowing through $A$, while an $f_A$ close to 0 suggests high porosity, permitting free flow of water.
    
    \item The drag coefficient $C_D$, which is a dimensionless factor that quantifies how hydrodynamic the flow-exposed shape of the UFD is, and affects the fluid's drag force exerted on the effective area $A'$.

    \item The ground clearance $z_{c}$, which is vertical distance from the floor to the lower chassis in vehicles. Both $f_A$ and $f_V$ may dramatically change when water depth exceeds this clearance height, such as the chassis height in vehicles.

    \item The friction coefficient $\mu$, which quantifies the ratio of frictional to normal forces between two contacting surfaces.
\end{enumerate}

To address the complexity of these parameters for UFDs, we compile a comprehensive inventory, presented and detailed in Appendix \ref{AppA}, for over 100 typified UFDs.
This inventory utilizes data from online catalogues for dimensions and mass.
For clearance height, an image-based manual identification method is employed.
We also adopt an image-based routine to estimate the effective drag area ($A'$).
This routine, applied to publicly available side images of UFDs (sources detailed in Supplementary Material), employs a binarization technique to extract the silhouette of the UFD.
The fraction $f_A$ is then calculated as the ratio between this silhouette and the flow-exposed area, under the assumption that the UFD is submerged to a specified water depth (Appendix \ref{AppA}).
For parameters like drag coefficient and submerged volume fraction, we either rely on empirically-derived values from existing literature or make simplifying geometric assumptions.

For UFDs of heterogeneous composition, our estimates of physical parameters are based on reasonable assumptions such as a range of densities and expected sizes, the latter observed in images from \cite{Bayon2023} (additional details and justification is included in Appendix \ref{AppA}).

\section{Mechanistic model}
\label{sec:MechModel}

\subsection{Limit states of stability}
\label{sec:limitstates}

\begin{figure}[H]
\includegraphics[width=7.5cm]{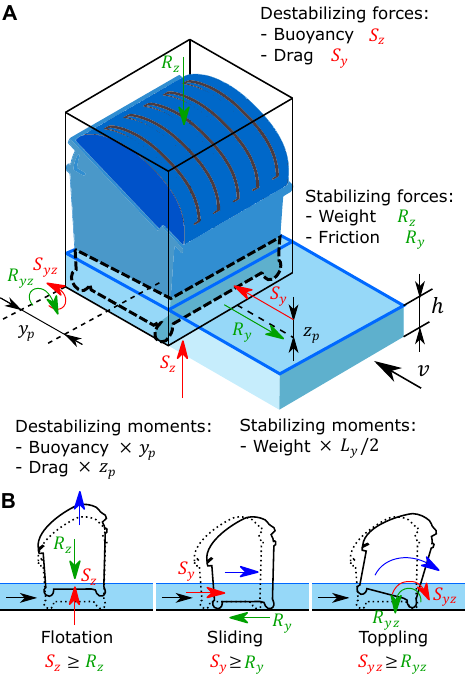}
\caption{Definition of (\textbf{A}) stabilizing and destabilizing forces and moments, and (\textbf{B}) limit states for the three destabilization modes.}
\label{Fig:destabilization}
\end{figure}

Aligned with previous studies \citep{Smith2019full, Martinez2020containers}, we establish a framework that considers the balance among four forces, 
drag ($S_y$) versus friction ($R_y$), 
buoyancy ($S_z$) versus weight ($R_z$),
and two moments --toppling load ($S_{yz}$) versus toppling resistance ($R_{yz}$), as illustrated in Figure \ref{Fig:destabilization} for a garbage bin, on the previously commonly identified UFDs.
These correspond to three different limit states of stability. 
These states are associated to the types of load ($S$) and resistance ($R$), as specified in Fig. \ref{Fig:destabilization}B: 
i) flotation ($S_z \geq R_z$, buoyant force exceeding weight), 
ii) sliding ($S_y \geq R_y$, drag force exceeding the UFD-floor friction) and 
iii) toppling ($S_{yz} \geq R_{yz}$, drag/buoyancy moment exceeding friction/weight moment).\\

\subsection{Flotation}

When buoyancy forces surpass the weight of an UFD, the object starts floating.
For this failure mode, the destabilising load ($S_z$) can be estimated as the buoyancy given by the Archimedes' principle, with $\rho_w$ being the water density and $g$ the gravity acceleration:

\begin{equation}
    S_z = \rho_w g f_V (h L_x L_y)
    \label{eq:S_z}
\end{equation}

The counteracting force or resisting component in the flotation limit state is the weight (mass $M = \rho_b L_x L_y L_z$, times gravity $g$):
\begin{equation}
    R_z = \rho_b L_x L_y L_z g 
    \label{eq:R_z}
\end{equation}

\subsection{Sliding}

Sliding occurs when the drag force, acting in the $y$-direction, destabilizes the UFD. 
This load can be estimated considering the UFD effective drag area ($A' = f_A L_x h$) and the drag coefficient ($C_D$):
\begin{equation}
    S_y = \frac{1}{2} \rho_w C_D f_A L_x h v^2
    \label{eq:S_y}
\end{equation}

We neglect the horizontal hydrostatic force as a destabilizing component. At low flow velocities, net hydrostatic forces are likely to cancel out due to compensation between the upstream and downstream sides. For high flow velocities and in supercritical conditions, the drag force is the predominant factor influencing the $y$-direction forces.

The resisting component in this limit state is given by the contact friction between the UFD and the floor, and can be expressed through the friction coefficient ($\mu$) as:
\begin{equation}
    R_y = \mu (R_z - S_z)
    \label{eq:R_y}
\end{equation}
which implies that, when the normal component cancels ($R_z \rightarrow S_z$),the horizontal as well ($R_y \rightarrow 0$). In other words, 

\subsection{Toppling}
Toppling may occur when the sum of destabilizing moments around an axis is larger than the sum of stabilizing moments opposing them. Taking the downstream rim of an UFD in the direction $x$ (Fig. \ref{Fig:destabilization}B) allows the estimation of moments:
\begin{equation}
    S_{yz} = S_y z_p + S_z y_p
    \label{eq:S_yz}
\end{equation}
with $z_p$ and $y_p$ as the lever arms at which forces $S_y$ and $S_z$ are applied, respectively (Fig. \ref{Fig:destabilization}A). These lever arms depend on the distribution of pressures over the wet surface. We make a justifiable approximation by considering:
\begin{equation}
    y_{p} \approx L_y /2
    \label{eq:y_p}
\end{equation}
and:
\begin{equation}
    z_{p} \approx
    \begin{cases}
      h/2 & \text{if } h \leq z_c\\
      z_c + (h-z_c)/2 & \text{if } h > z_c\\ 
    \end{cases}
    \label{eq:y_p}
\end{equation}
with the $z_p$ definition acknowledging that most of the flow impacts the UFD over $z_c$, once it is submerged over the clearance height ($h>z_c$).

The toppling resistance will be given by:
\begin{equation}
    R_{yz} = R_z \, L_y/2
    \label{eq:R_yz}
\end{equation}

\section{Probabilistic framework} 
\label{sec:MC_uncertainty_def}

\subsection{General remarks}

Utilizing the classification scheme introduced by \cite{Bayon2023}, as outlined in Table \ref{tab:Classification_UFDs}, we categorize the properties of UFDs into three main groups: i) Vehicles (UFD-V), ii) Furniture (UFD-F), and iii) Heterogeneous UFD (UFD-H). We also incorporate all the subgroups presented in Table \ref{tab:Classification_UFDs}.
This categorization is useful because it allows us to define ranges of realistic values for all parameters used in the mechanistic model discussed in Section \ref{sec:MechModel}.

To identify realistic parameters for the UFDs, serving as inputs for the mechanistic model detailed in Section \ref{sec:MechModel}, we conduct a comprehensive examination. This examination draws upon various sources, contingent on the specific model of the typified UFDs under investigation.
The survey of key properties -- covering those identified in Section \ref{sec:UFD_def}-- is further elaborated in Appendix \ref{AppA} and Supplementary Material, where all sources are included.
For heterogeneous UFDs, we base our parametrization on a comprehensive literature review, complemented by justified boundaries.

\begin{figure}[H]
\includegraphics[width=7.5cm]{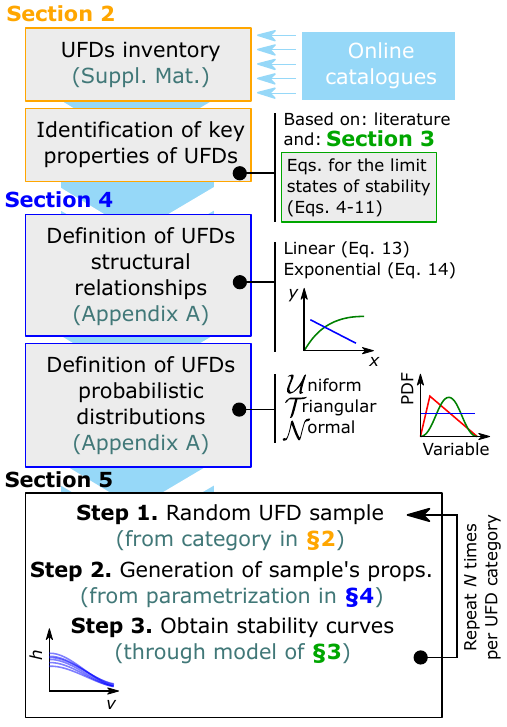}
\caption{Monte Carlo approach to the estimation of stability curves for different UFDs.}
\label{Fig:workflow}
\end{figure}

In the following, we explain how we leverage this key properties inventory for UFDs and define functional relationships and probabilistic density functions (Section \ref{sec:probabilistic_framework} and Appendix \ref{AppA}). These serve to capture the diversity inherent to each category of UFDs. The ultimate goal is to generate random, yet realistic, samples from a UFD subcategory for the purpose of performing Monte Carlo analyses of their stability, as demonstrated through the mechanistic model in Section \ref{sec:MechModel}. This approach is illustrated in Figure \ref{Fig:workflow}.


\subsection{Structural relationships for UFDs' key variables and their variability} 
\label{sec:probabilistic_framework}

Each UFD, whether it be a car, a trash bin, or another object, exhibits distinct characteristics, a variability that our model aims to capture.
Certain characteristics, such as dimensions $L_x$ and $L_y$ in vehicles, often scale together —-these are identified as structural relationships.
These co-dependencies across multiple variables need to be considered, to ensure that, when we generate a synthetic UFD to study its stability, we are generating a realistic UFD.

In contrast, other UFD properties might exhibit randomness within defined limits and are categorized as randomistic. For such properties, we utilize Probability Density Functions (PDFs).
The choice of PDF for each variable is based on an empirical evaluation of the Cumulative Density Function, typically supplemented by physical reasoning. For example, if a characteristic must be strictly positive, a $\mathcal{U}$ (uniform distribution) or $\mathcal{T}$ (triangular distribution) is favored over a $\mathcal{N}$ (Normal distribution), which carries a non-zero probability of producing negative values.

Either structural or randomistic, the characteristics of each UFD subcategory can be explained through the following equation:

\begin{equation}
    {y(x)} = \Upsilon(x) + y'
    \label{eq:correlation}
\end{equation}

Here, $x$ is the dependent variable, while $y$ is the independent variable.
$\Upsilon(x)$ is a function that portrays the trend --or structural relationship-- of $y$ relative to $x$, while $y'$ denotes a stochastic component that introduces additional variability into $y$. 
The functional form of $\Upsilon(x)$ is flexible, but we primarily investigate two functional relationships—linear and saturation curve—besides a null function:

\begin{equation}
    \Upsilon =  a \,x + b
    \label{eq:Linear}
\end{equation}
\begin{equation}    \Upsilon =  a \left( 1 - \exp \left( b \, x\right)\right)
    \label{eq:OneMinusExp}
\end{equation}

The parameters $a$ and $b$ are optimized through error minimization algorithms to best fit the observed UFD data, as described in Appendix \ref{AppA}. For certain co-dependencies, Eq. \ref{eq:OneMinusExp} offers a superior fit. For example, within all vehicles subcategories (UFD-V), an increase in $L_x$ is accompanied by an increase in $L_y$, but only asymptotically up to \mbox{2.50 m}, as constrained by regulations such as the European Union Council Directive 96/53/EC.

When no correlation is observed among pairs of variables, $\Upsilon(x) \equiv 0$ and only the randomistic component ($y'$) in Eq. \ref{eq:correlation} defines an UFD characteristic ($y$). This is the dominant situation in UFD-H, and incorporates further variability (uncertainty) into the stability curves.
For $y'$, we consider three potential PDFs:  $\mathcal{U}$, $\mathcal{T}$, or $\mathcal{N}$. 

A comprehensive review of the functional relationships for $\Upsilon(x)$ and the PDFs for $y'$ across all UFD subcategories is provided in Appendix \ref{AppA}. All relevant data and analysis codes are available in the Electronic Supplementary Material.

\subsection{Monte Carlo analysis}
\label{sec:MC_analysis}

A Monte Carlo analysis is performed to probabilistically estimate the stability curves for different UFD categories, as outlined in Table \ref{tab:Classification_UFDs}. This analysis is building upon the mechanistic model of Section \ref{sec:MechModel} (Fig. \ref{Fig:destabilization}) and the input parametrization defined across Section \ref{sec:MC_uncertainty_def} and Appendix \ref{AppA}. 
The procedure for the analysis is conducted for each UFD subcategory and follows the protocol here described (Fig. \ref{Fig:workflow}):
\begin{enumerate}
    \item We generate 1,000 different samples from a given UFD subcategory (e.g., UFD-F2), based on structural and stochastic characteristics of each group derived in Section \ref{sec:probabilistic_framework}, thereby simulating a realistic range of physical properties.
    \item Stability limit state equations —for flotation, sliding, and toppling— are solved for each synthetic UFD generated.
    \item For each synthetic UFD, the minimum depth $h$ required to initiate motion at a given velocity $v$ is identified by considering all three instability modes and is designated as the value for the stability curve.
    \item Results are visualized 
    and data is stored by subcategory in the Electronic Supplementary Material.
\end{enumerate}

\section{Results} 

\subsection{Application and verification of the mechanistic framework}
\label{sec:calibration}

\begin{figure}[H]
\centering
\includegraphics[width=0.85\textwidth]{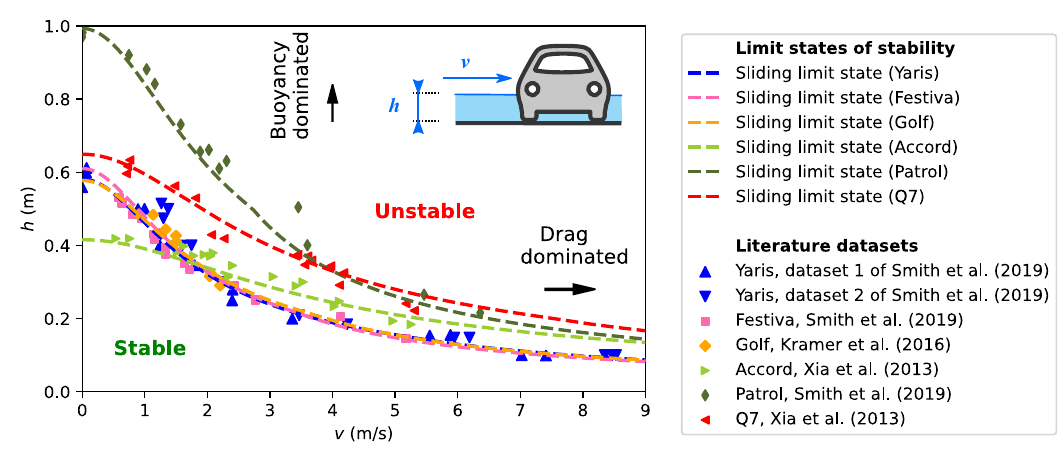}
\caption{Stability curves obtained with the mechanistic model (Section \ref{sec:MechModel}) for specific car models tested in laboratory in literature \citep{Xia2014criterion, Kramer2016safety, Smith2019full}.}
\label{Fig:Model_verif}
\end{figure}

In this section, we utilize various case studies from literature to evaluate the accuracy and applicability of the proposed mechanistic model proposed in Section \ref{sec:MechModel}. We focus the verification study on cars (UFD-V2, Table \ref{tab:Classification_UFDs}), owing to the wealth of available research data.  Figure \ref{Fig:Model_verif} presents both the experimental data from literature and the predicted limit states of our mechanistic model (Section \ref{sec:MechModel}).

\begin{table}[H]
    \centering
    \begin{tabular}{ p{0.7cm} p{0.7cm} p{0.7cm} p{0.7cm} p{0.7cm} p{0.7cm} p{0.7cm} p{0.7cm} p{0.7cm} p{0.7cm} p{1.0cm} p{2.7cm}}
        \hline
        Car        & $M$ (kg) & $L_x$ (m) & $L_y$ (m) & $L_z$ (m) & $z_c$ (m) & $C_D$ (-) & $f_{V1}$ (-) & $f_{V2}$ (-) &
        $\mu$ (-) & Exp. scale & Source\\
        \hline
        Yaris & 1045 & 4.30 & 1.69 & 1.46 & 0.16 & 1.38 & 0.05 & 0.32*& 0.30 & 1:1 & \cite{Smith2019full}  \\
        Festiva & 790 & 3.62 & 1.61 & 1.46 & 0.22 & 1.38 & 0.05 & 0.32*& 0.30 & 1:1 & \cite{Smith2019full}  \\
        Golf & 1261 & 4.26 & 1.80 & 1.45 & 0.21 & 1.70* & 0.05 & 0.42* & 0.30 & 1:1 & \cite{Kramer2016safety}  \\
        Accord & 1615 & 4.95 & 1.85 & 1.48 & 0.12 & 0.75* & 0.05 & 0.58* & 0.30 & 1:14/1:24  & \cite{Xia2014criterion} \\
        Patrol & 2478 & 4.97 & 1.84 & 1.94 & 0.50 & 1.38 & 0.05 & 0.50*& 0.30 & 1:1 & \cite{Smith2019full}  \\
        Q7 & 2345 & 5.09 & 1.98 & 1.74 & 0.20 & 0.85* & 0.05 & 0.50* & 0.30 & 1:14/1:24  & \cite{Xia2014criterion} \\
        \hline
    \end{tabular}
    \caption{Key characteristics of the UFD-V2 models used for verification. The $f_A$ used is that parameterized in Appendix \ref{AppA} for UFD-V2, Eq. \ref{eq:fA_poly}, with stochastic term $f_A'$ = 0. $f_{V1}$ corresponds to the submerged volume contribution of the wheels, assumed constant, and $f_{V2}$ corresponds to the above-clearance contribution. Values with (*) correspond to best fit (calibrated), which is only conducted for missing data.}
    \label{tab:verification_cars}
\end{table}

Some parameters ($M$, $L_x$, $L_y$, $L_z$, $z_c$), necessary for the application of the model (Eqs. \ref{eq:S_z} to \ref{eq:R_yz}), are directly incorporated from the original case studies.
In all instances, we consider the friction coefficient ($\mu$) as 0.3, in line with reporting from laboratory experiments \citep[see study of][]{Smith2019full}.

For the evaluation of $f_A$ at each depth, we use the data-driven polynomial approximation specified for UFD-V2 in Appendix \ref{AppA} (Eq. \ref{eq:fA_poly}, excluding the stochastic term $f_A'$).
Unspecified parameters such as $f_{V,2}$ and, occasionally, $C_D$ are calibrated iteratively. Given that $C_D$ impacts stability at high flow velocities, while $f_{V,2}$ is influential at low velocities, these parameters can be calibrated independently. 
We recognize that the drag coefficient ($C_D$) varies among the considered case studies. This variation is reasonable and can be justified on geometrical differences in shapes of car models, given that \cite{Xia2014criterion}.

Inspection of Figure \ref{Fig:Model_verif} reveals that the stability limit state curves generated by our model, in conjunction with physical properties, are physically consistent and describe the laboratory data in good detail. At near-zero velocities, flotation becomes the governing instability mode. For non-zero velocities, buoyancy lowers the threshold for sliding and toppling by reducing the effective normal force ($R_y - S_y$, see Equations \ref{eq:R_y} and \ref{eq:S_yz}). The stability curves exhibit a notable flattening as water levels drop below the clearance height ($z_c$), which corresponds to a diminishing effect of $f_A$ and $f_V$. In all our verification cases, toppling proves to be a less likely failure mode than either sliding or flotation, occurring at relatively lower water depths.

\subsection{Probabilistic assessment of stability curves}
\label{sec:prob_results}

\begin{figure}[H]
\centering
\includegraphics[width=1.\textwidth]{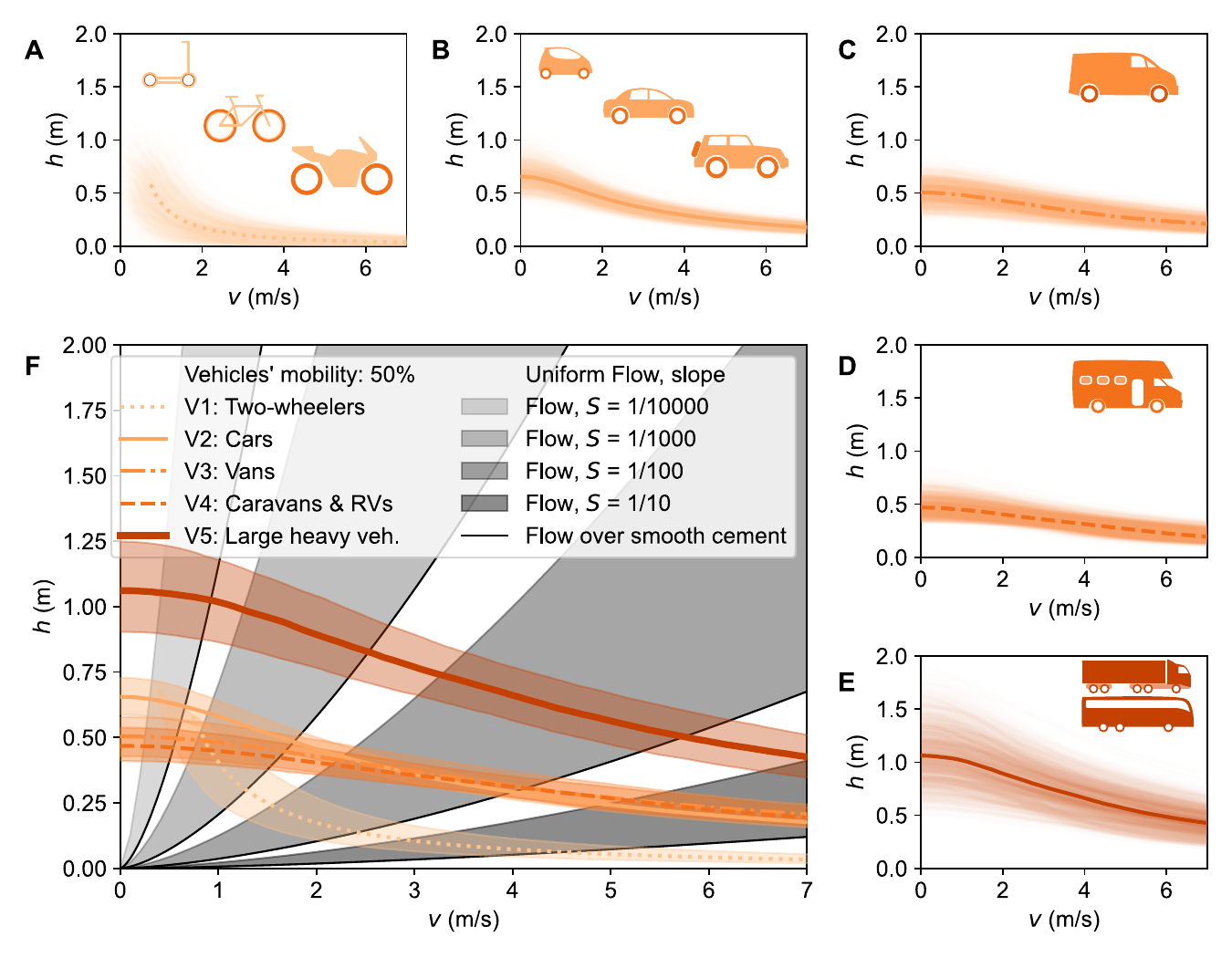}
\caption{Stability curves for vehicles: \textbf{A} Two-wheelers (UFD-V1), \textbf{B} Cars (UFD-V2), \textbf{C} vans (UFD-V3), \textbf{D} caravans \& RVs (UFD-V4), \textbf{E} heavy vehicles (UFD-V5). \textbf{F} Vehicles stability curves intersection with uniform flow over different urban surfaces \citep[from smooth cement to excavated gravel,][]{Chow1959} at different slopes.}
\label{Fig:Stability_vehicles_complete}
\end{figure}

\begin{figure}[H]
\centering
\includegraphics[width=1.\textwidth]{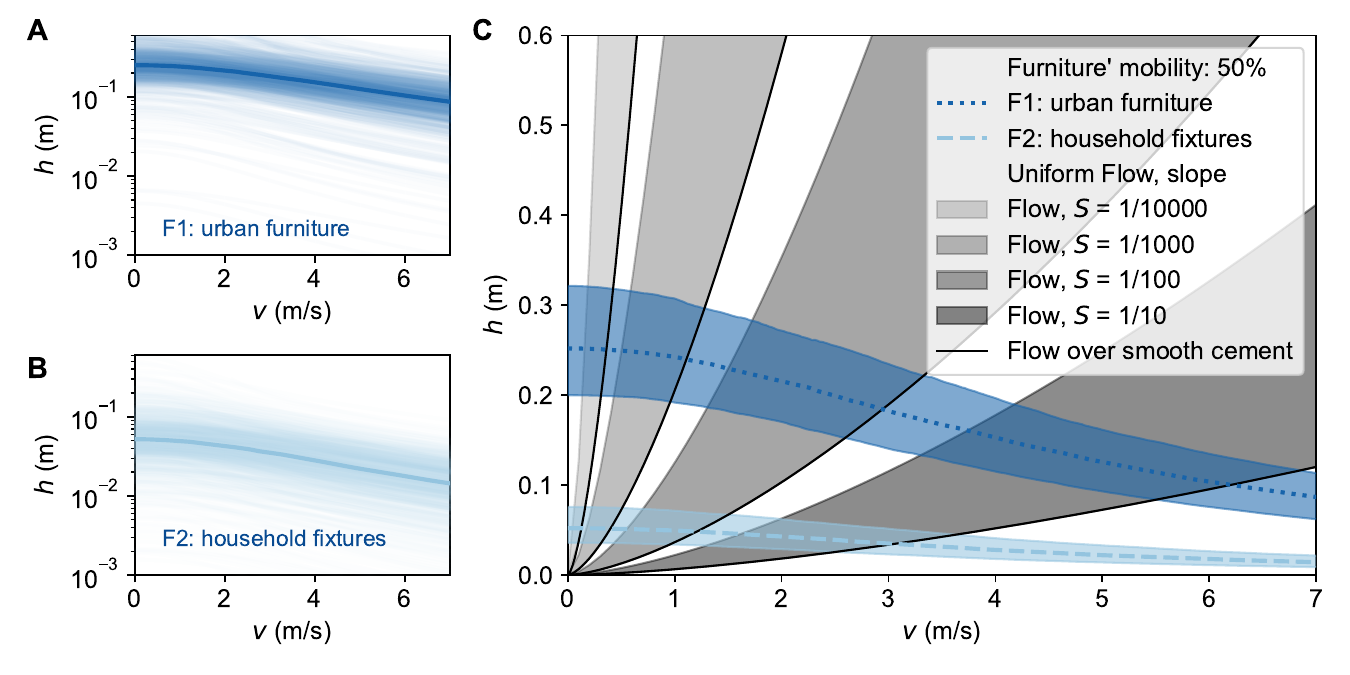}
\caption{Stability curves for furniture: \textbf{A} urban furniture (UFD-F1), \textbf{B} private furniture (UFD-F2). \textbf{C} Furniture stability curves intersection with uniform flow over different urban surfaces \citep[from smooth cement to excavated gravel,][]{Chow1959} at different slopes.}
\label{Fig:Stability_furniture_complete}
\end{figure}

\begin{figure}[H]
\centering
\includegraphics[width=1.\textwidth]{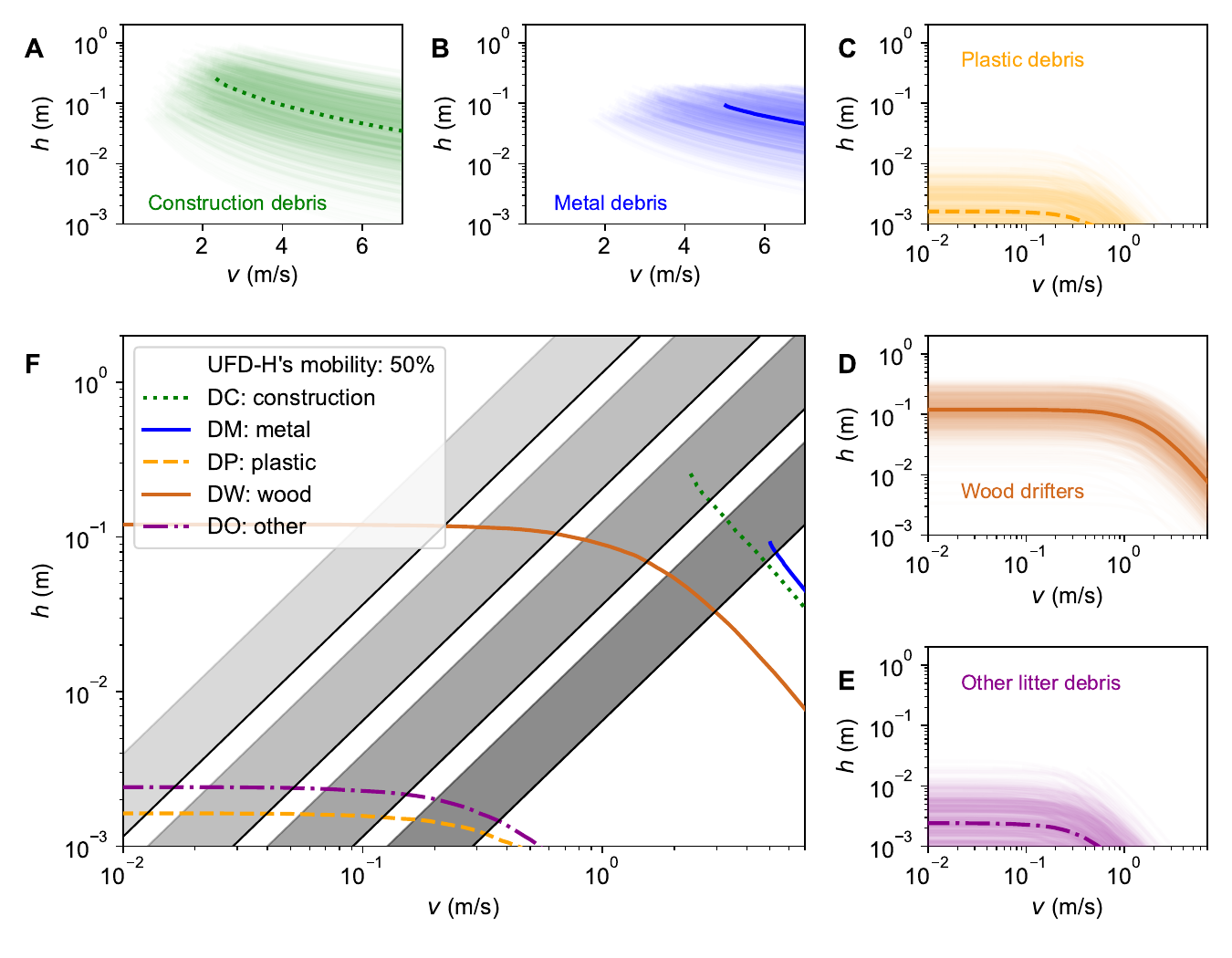}
\caption{Stability curves for UFD-H: \textbf{A} construction debris (UFD-DC), \textbf{B} metal debris (UFD-DM), \textbf{C} plastic debris (UFD-DP), \textbf{D} wood (UFD-DW), \textbf{E} other debris (UFD-DO). \textbf{F} Heterogeneous UFD curves intersection with uniform flows over different urban surfaces at different slopes (same as in Fig. \ref{Fig:Stability_vehicles_complete}) \citep[from smooth cement to excavated gravel,][]{Chow1959} at different slopes.}
\label{Fig:Stability_debris_complete}
\end{figure}

In this subsection, we extend the application of the mechanistic model to encompass all UFD subcategories, deploying the previously outlined Monte Carlo simulation approach (see Fig. \ref{Fig:workflow}) that follows the inherent variability of UFDs. For clarity, Figs. \ref{Fig:Stability_vehicles_complete}F,  \ref{Fig:Stability_furniture_complete}C and Fig \ref{Fig:Stability_debris_complete}F incorporate flow conditions based on uniform flow estimations, which are based on the Manning formula \citep{Chow1959}. This enables the estimation of pairs of $h$-$v$ for a given slope and roughness coefficient. These highlighted regions within the plots intersect the stability curves, thereby identifying the conditions under which UFD mobility is likely to be initiated. For that, we consider a wide array of slopes and surface roughness, ranging from smooth cement to gravel bed, with corresponding Manning numbers $n = 0.011$ to $0.025$. These are representative of reasonable surfaces and slopes found in urban setups. The highest value of the Manning coefficient ($n = 0.025$) is also representative of short grass.


Our probabilistic analysis of vehicles' stability reveals several critical insights (Figure \ref{Fig:Stability_vehicles_complete}):

\begin{itemize}
    \item Two-wheelers (UFD-V1) are generally less stable than other vehicle types. However, some V1 vehicles possess higher bulk density than water (due to small internal volumes, metal construction and no cabin). This results in a threshold velocity below which they remain unconditionally stable; i.e., sinking regardless of the flow depth. 
    \item Cars (UFD-V2) exhibit stability conditions very similar to those seen in the verification cases, with the dispersion of the data aligning with the variability observed in the verification cases. This reaffirms that this type of UFD is well represented in the literature.
    \item Vans (UFD-V3), caravans and RVs (UFD-V4) demonstrate a behavior comparable to UFD-V2, although caravans exhibit a considerably narrower distribution of stability curves. This is because, with increasing size, bulk density tends to reduce for habitable vehicles. Therefore, both vans and caravans tend to be less stable than cars on flatter slopes when flotation becomes the dominant mode of failure. This early onset of mobility, coupled with their larger volumes, increases flood hazard due to a higher potential for infrastructure clogging and damage. Given that these larger vehicles are commonly observed in flood events \citep{Bayon2023}, mitigating flood hazards would require additional focus on UFD-V3 and UFD-V4. 
    \item Heavy vehicles (UFD-V5), owing to their massive scale, exhibit substantially higher stability. However, this category exhibits a wider dispersion due to the considerable range of properties; for instance, $L_x$ ranges from 6 to 18 m. Our findings suggest that these vehicles could contribute to flood hazards under flood situations with depths exceeding 1 m, particularly when also exposed to high velocities ($v >$ 2 m/s). UFD-V5, with masses one to two orders of magnitude greater than other UFD-Vs, may substantially amplify potential infrastructure damage due to impacts \citep{Zhang2018debris, Jalayer2018debris}, in addition to the also increased flood-infrastructure clogging hazard.
\end{itemize}


When considering furniture (UFD-F), it is discernible that both UFD-F1 and UFD-F2 are highly movable. Urban fixtures (UFD-F1) become unstable at approximate depths of 25 cm, while household equipment (UFD-F2) exhibits instability at depths significantly below 10 cm. This signals an earlier onset of transport as compared to vehicles. Nonetheless, UFD-F2 are commonly installed within private properties, potentially attenuating the direct impact of flooding.

UFD-F, frequently visible during and post flood episodes (see Figure \ref{Fig:UFDs_pics}), constitutes approximately 7.4\% of UFDs relative frequency \citep{Bayon2023}. This statistic, together with the disclosed early mobility, suggests that the role of furniture in urban floods might have been consistently underestimated.

For heterogeneous debris and drifters (UFD-H), our analysis (Figure \ref{Fig:Stability_debris_complete}) indicates that stability characteristics are closely related to the material density. Plastic debris is the least stable, with the vast majority of items becoming mobilized at water depths of below a few centimeters. This is followed closely by other litter, which shows a very similar behaviour.
Wood remains considerably more stable up to flood depths of 10 cm for low velocities (< 1 m/s). Velocities and depths observed are consistent with the wood stability study of \cite{Braudrick2000logs}. The most stable types within the UFD-H category are construction materials and metal, which require exceptionally large velocities for the inception of mobility -—of at least 3 to 5 m/s—- typically becoming dislodged only after impacts from other UFDs (such as trucks or cars) with infrastructure or when located near construction sites adjacent to streets.

\section{Discussion: sequential mobilization of UFDs}
\label{sec:discussion}

\begin{figure}[H]
\includegraphics[width=8cm]{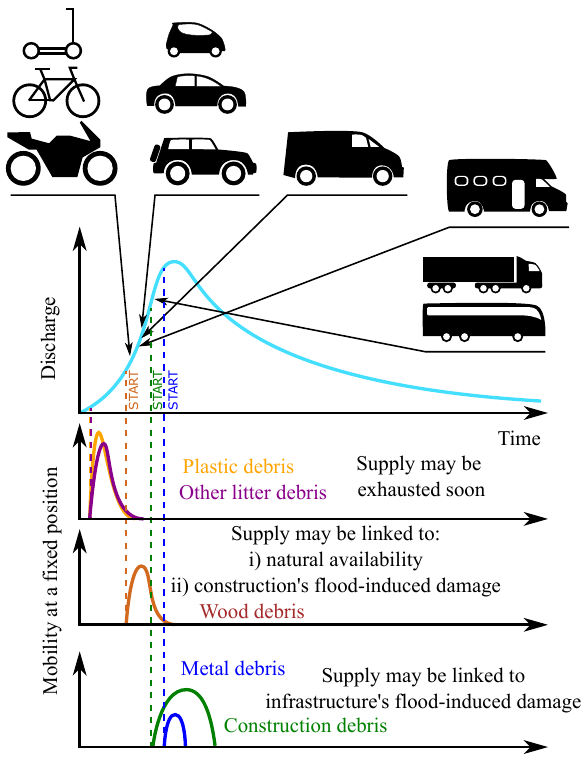}
\caption{Complete UFD mobility at varying flood hydrograph stages, based on probabilistic onset calculations of Figs. \ref{Fig:Stability_vehicles_complete}, \ref{Fig:Stability_furniture_complete} and \ref{Fig:Stability_debris_complete}.}
\label{Fig:Debris_hydrograph}
\end{figure}

Our study sheds light on the mobility of UFD, enabling inferences regarding their sequential mobilization during a flood event.
Figures \ref{Fig:Stability_vehicles_complete}, \ref{Fig:Stability_furniture_complete} and \ref{Fig:Stability_debris_complete} presented the $h-v$ pairs for which a given UFD is mobilised. Assuming a low slope configuration, we can extract and order which UFDs are mobilised. Assuming that a limited supply of UFDs is available, this allows sketching the expected UFD-graphs presented in Fig. \ref{Fig:Debris_hydrograph}.
First to mobilize is typically plastic debris, which is highly mobile and remains so until supply exhaustion (Figure \ref{Fig:Debris_hydrograph}). This aligns with previous hydrological observations on plastic pollution \citep{VanEmmerik2019seasonality, VanEmmerik2019Seine, VanEmmerik2022river, Bayon2023} and mechanistic studies \citep{Goral2023settling, VanEmmerik2023stability}, also including those focusing on early transport \citep{Kuizenga2022macro, Valero2022surfaced, Lofty2023bedload} that show mild settling velocities.

Next to mobilize are loose wooden pieces, more specifically, those that are buoyant. 
During the falling limb of the hydrograph, denser wood may settle sooner, while buoyant pieces may continue to travel farther, provided they are not obstructed, as long as the water depth remains large enough. This is also consistent with observations of \cite{Ruiz2014twodimensional} and \cite{Ruiz2016hydrograph}.

The behavior of UFD-F during a flood is more nuanced. The stability curves obtained (Figure \ref{Fig:Stability_furniture_complete}) suggest early mobilization, similar to smaller wood debris and other highly-mobile UFD-H. However, some UFD-F may remain anchored or weighted down (for example, by a water tank or a trash bin), as suggested by  \cite{Martinez2020containers}, thus introducing variability in their mobilization patterns, which might deviate from our predictions. Large, loose containers may float at shallow water depths, exhibiting behaviours similar to barges. Household equipment (UFD-F2), despite being highly mobile, may remain sheltered as they are frequently situated in backyards or indoor areas.

As flood depths and velocities increase, vehicles begin to mobilize, initially with two-wheelers (UFD-V1) on steep slopes. In areas with gentle slopes, the velocity may not reach the required threshold for mobilization (Figure \ref{Fig:Stability_vehicles_complete}A). As the flood progresses towards the peak, two scenarios may unfold: in low-slope areas (low $v$, high $h$), vans (UFD-V3), caravans and RVs (UFD-V4) may destabilize more easily than cars, whereas the opposite scenario is expected for high-slope urban areas (Figure \ref{Fig:Stability_vehicles_complete}B-D,F). As flow intensifies, larger vehicles like trucks (UFD-V5) may also eventually be mobilized, assuming water depths clearly exceed one meter.

Lastly, construction and metal debris are typically mobilized at high velocities and often occur after other UFDs have caused infrastructural damage  \citep{Jalayer2018debris, Zhang2018debris}. 
Even when fully submerged, very high velocities are needed to transport these debris (Fig. \ref{Fig:Stability_debris_complete}A,B) and are likely to settle as soon as floodwaters begin to recede.

\section{Conclusions}  
\label{sec:conclusions}

This research provides a comprehensive analysis of the mobility of a wide range of Urban Flood Drifters (UFDs), a significant but under-studied contributor to flood risk  \citep{Dewals2021july, Mohr2022multi, Bayon2023}. In contrast to prior work focused mainly on vehicle stability during floods, we offer a broader assessment of UFD stability. Our contributions are as follows:
\begin{enumerate}
    \item We developed a mechanistic model addressing the limit states of flotation, sliding, and toppling (Section \ref{sec:MechModel}, Equations \ref{eq:S_z}--\ref{eq:R_yz}). This model has been validated against existing vehicle stability studies, demonstrating robust predictive capabilities (Fig. \ref{Fig:Model_verif}).
    \item We established a Monte Carlo-based probabilistic model (Section \ref{sec:MC_uncertainty_def}) that accommodates the variability and uncertainty in UFD characteristics.
    \item A detailed inventory of UFD physical properties was compiled, serving as essential inputs for the developed models. These properties align with the UFD categories identified by \cite{Bayon2023} (Table \ref{tab:Classification_UFDs}).
    \item Probability distributions for UFD characteristics were defined, grounded in empirical data (Eq. \ref{eq:correlation}, Appendices \ref{AppA}).    
    \item Using the above frameworks, we conducted Monte Carlo analyses to generate limit state curves for a wide array of UFDs (Figs. \ref{Fig:Stability_vehicles_complete}, \ref{Fig:Stability_furniture_complete}, \ref{Fig:Stability_debris_complete}).
    \item We delineated the sequence of UFD mobilization during urban flooding, offering insights into the transport dynamics of various UFDs, ranging from plastics to heavy vehicles and construction debris (Fig. \ref{Fig:Debris_hydrograph}).
\end{enumerate}

The derived stability curves, available in the Electronic Supplemental Material, can be integrated into 2D flood models. Coupled with source/supply estimations, these curves enable, for the first time, the prediction of UFD travel patterns during urban flooding. This aids in identifying high-risk zones for clogging and backwater effects, thereby enhancing urban flood risk mitigation strategies.








\appendixfigures  

\appendixtables   


\appendix
\section{Characterization and uncertainties in typified UFDs' key properties} 
\label{AppA}

This appendix explains in detail the parametrization of the variables that determine the stability of Urban Flood Drifters (UFDs), as outlined in Section \ref{sec:MechModel}. These variables are addressed one-by-one and by UFD subcategory (c.f. Table \ref{tab:Classification_UFDs}), thus precisely tailoring the parametrization (see Eq. \ref{eq:correlation}) to the distinct variability and uncertainties inherent to each category. The parametrization of Eq. \ref{eq:correlation} admits two terms for each dependent variable (e.g., $y(x)$): the structural function ($\Upsilon(x)$) and the stochastic component ($y'$). For the structural dependency, two functions are considered (besides the null function): linear \mbox{(Eq. \ref{eq:Linear})} and saturation \mbox{(Eq. \ref{eq:OneMinusExp}).} For the stochastic component, the probability density functions considered are: $\mathcal{C}$ (Constant, single value),  $\mathcal{U}$ (Uniform), $\mathcal{T}$ (Triangular) and $\mathcal{N}$ (Normal); whose parameters are defined in this Appendix.

The parametrization process is systematically organized for each variable as follows:
\begin{itemize}
    \item Dimensions of the bounding box ($L_x$, $L_y$ and $L_z$), in Section \ref{sec:length-scales}.
    \item Clearance height ($z_c$) or mass ($M$), in Section \ref{sec:clearance}.
    \item Bulk density ($\rho_b$), in Section \ref{sec:rho_b},
    \item Fraction of effective drag area ($f_A$), in Section \ref{sec:fa}.
    \item Fraction of submerged volume ($f_V$), in Section \ref{sec:fV_formula}.
    \item Friction coefficient ($\mu$), in Section \ref{sec:mu}.
    \item Drag coefficient ($C_D$), in Section \ref{sec:CD_param}.
\end{itemize}
 
\subsection{Dimensions of the bounding box ($L_x$, $L_y$ and $L_z$) and their co-dependency}
\label{sec:length-scales}

The bounding box of an UFD is defined by three length scales: $L_x$ (largest horizontal length), $L_y$ (shortest horizontal length), and $L_z$ (height), as depicted in Figure \ref{Fig:BoundingBox_fA_fV}A-C.

In this analysis, $L_x$ is selected as the independent random variable for UFD-Vs, with its variability serving as the foundation for deriving other physical characteristics.
In other words, when running the Monte Carlo procedure (Fig. \ref{Fig:workflow}), the first step done to generate a synthetic UFD sample is to generate a $L_x$ value, which is governed by the primary descriptors presented in 
\mbox{Table \ref{tab:L_x_distro-vehicles}.}

\begin{table}[H]
    \centering
    \begin{tabular}{p{0.5cm} p{0.5cm} p{0.5cm} p{0.5cm}   p{0.75cm} p{1.25cm} p{1.25cm} p{1.25cm}}
        \hline
        ID &  $y$ & $x$ & $\Upsilon$ (fit) & $y'$ (PDF) &  min($y'$) & mode($y'$) & max($y'$) \\
        \hline
        V1 & $L_x$ & - & - & $\mathcal{T}$ & 1.0   &   1.8   &   2.6 \\
        V2 & $L_x$ & - & -   &   $\mathcal{T}$ & 2.7   &   4.4   &   5.2\\
        V3 & $L_x$ & - & -   &   $\mathcal{U}$ & 4.0   &   {-}   &   7.5\\
        V4 & $L_x$ & - & - &   $\mathcal{U}$ & 4.9   &   {-}   &   8.0 \\
        V5 & $L_x$  & - & -  &  $\mathcal{T}$ & 6.0   &   9.0   &   18.0* \\
        \hline
        F1 & $L_x$ & - & - &  $\mathcal{U}$ & 0.30   &   {-}   &   6.0 \\
        F2 & $L_x$ & - & - &  $\mathcal{T}$ & 0.40   &   2.25   &   3.4  \\
        \hline
    \end{tabular}
    \caption{Components of Eq. \ref{eq:correlation} for the description of the length $L_x$ for UFD-V and UFD-F. Probabilistic distributions based on observations presented in Figure \ref{Fig:SimpleGeom-vehicles-L}A,B. Min: minimum value of the variable ($y$) represented by the probabilistic distribution, Mode: mode (only for triangular), Max: maximum value.
    (*) This upper bound corresponds to the length of an articulated bus \citep[see Table 5.3 of][]{Vuchic2007urban}.}
    \label{tab:L_x_distro-vehicles}
\end{table}

\begin{figure}[H]
\centering
\includegraphics[width=1.0\textwidth]{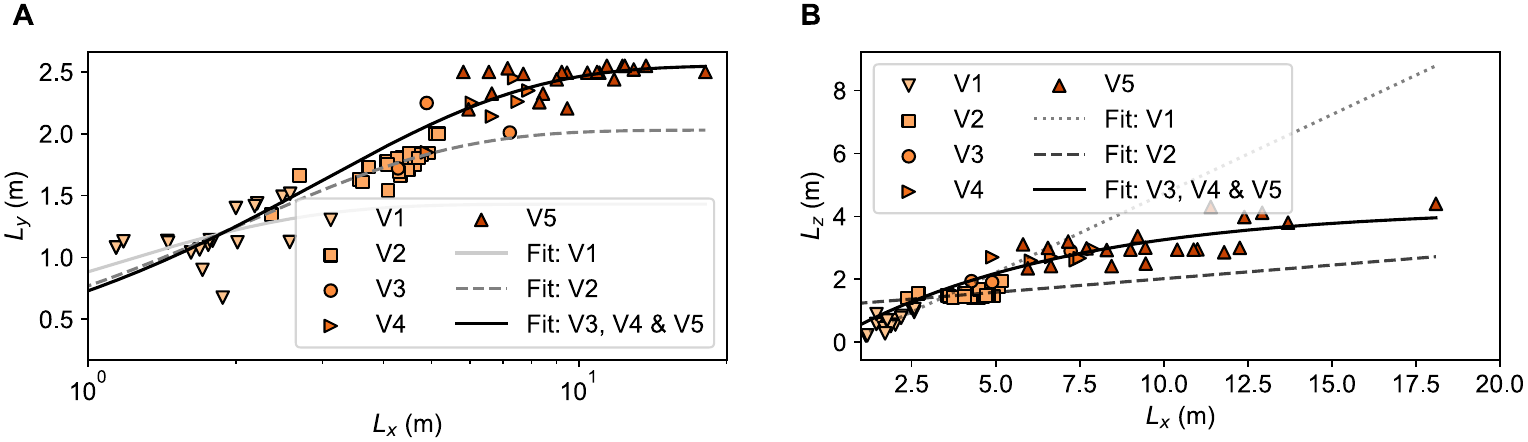}
\caption{Co-dependencies between (A) length and width and (B) length and height in vehicles (UFD-V1 to UFD-V5).}
\label{Fig:SimpleGeom-vehicles-L}
\end{figure}

Figure \ref{Fig:SimpleGeom-vehicles-L} illustrates the codependency of the geometrical properties of UFD-V ($L_x$, $L_y$, $L_z$). A clear correlation is noted between $L_y$ and $L_x$ (Fig. \ref{Fig:SimpleGeom-vehicles-L}A), and between $L_z$ and $L_x$ (Fig. \ref{Fig:SimpleGeom-vehicles-L}B). With an increase in vehicle length $L_x$, all other dimensions tend to grow until a limit is reached at approximately 2.5 m for $L_y$ and around 4 m for $L_z$. These limits may be attributed to road regulation dimensions and potential clearance constraints for bridges.

Based on these observations, separate functional relationships (Eq. \ref{eq:correlation}) are produced for $L_y$ and $L_z$ for V1, V2 and V3-V5 (Table \ref{tab:Ly-a-b-RMSE-vehicle_L}, and also presented in Figure \ref{Fig:SimpleGeom-vehicles-L}A,B).

\begin{table}[H]
    \centering
    \begin{tabular}{ p{1cm} p{0.5cm} p{1.0cm} p{1.25cm} p{2.2cm} p{2.2cm} p{2.2cm} p{0.75cm} }
        \hline
        ID              & $y$ & $x$ & $\Upsilon$ (fit) &  $a$  & $b$ & RMSE & $y'$ (PDF)\\
        \hline
        V1              & $L_y$ & {$L_x$} &  Saturation    & 1.431 m & 0.963 m$^{-1}$ & 0.182 m &  $\mathcal{U}$\\
        {}              & $L_z$ & {$L_x$} & Linear & 0.502 m/m & -0.295 m & 0.151 m &  $\mathcal{U}$\\
        V2              & $L_y$ &  {$L_x$} & Saturation & 2.030 m & 0.476 m$^{-1}$ & 0.088 m &  $\mathcal{N}$\\
        {}              & $L_z$ &  {$L_x$} & Linear & 0.086 m/m & 1.156 m  & 0.117 m &  $\mathcal{N}$\\
        V3, V4, \& V5   & $L_y$ &  {$L_x$} & Saturation     & 2.551 m & 0.337 m$^{-1}$  & 0.137 m &  $\mathcal{N}$\\
        {}              & $L_z$ &  {$L_x$} & Saturation & 4.268 m & 0.144 m$^{-1}$  & 0.396 m &  $\mathcal{N}$\\
        \hline
        F1  & $L_y$ & {$L_x$} &  Saturation    & 2.339 m & 0.479 m$^{-1}$ & 0.253 m &  $\mathcal{N}$\\
        {}  & $L_z$ & {$L_y$} &  Linear    & 0.476 m/m & 0.648 m & 0.513 m &  $\mathcal{U}$\\
        F2  & $L_y$ & {$L_x$} &  Linear  & 0.863 m/m & -0.127 m & 0.252 m &  $\mathcal{N}$\\
        {}  & $L_z$ & {$L_y$} &  Saturation    & 2.727 m & 0.628 m$^{-1}$ & 0.389 m &  $\mathcal{U}$\\
        \hline
    \end{tabular}
    \caption{Components of Eq. \ref{eq:correlation} for the description of the width ($L_y$) and height ($L_z$) of vehicles (UFD-V) and furniture (UFD-F). Fits shown in Figure \ref{Fig:SimpleGeom-vehicles-L}A,B, Figure \ref{Fig:SimpleGeom-furniture_L}. $\mathcal{N}$ distributions are zero-mean and standard deviation is assumed to be the data fit RMSE. $\mathcal{U}$ distributions are zero-mean and RMSE-based (min and max limits setup to $+/-\sqrt{3}$ RMSE, respectively, to satisfy STD($\mathcal{U}$) = RMSE).}
    \label{tab:Ly-a-b-RMSE-vehicle_L}
\end{table}

\begin{figure}[H]
\centering
\includegraphics[width=1\textwidth]{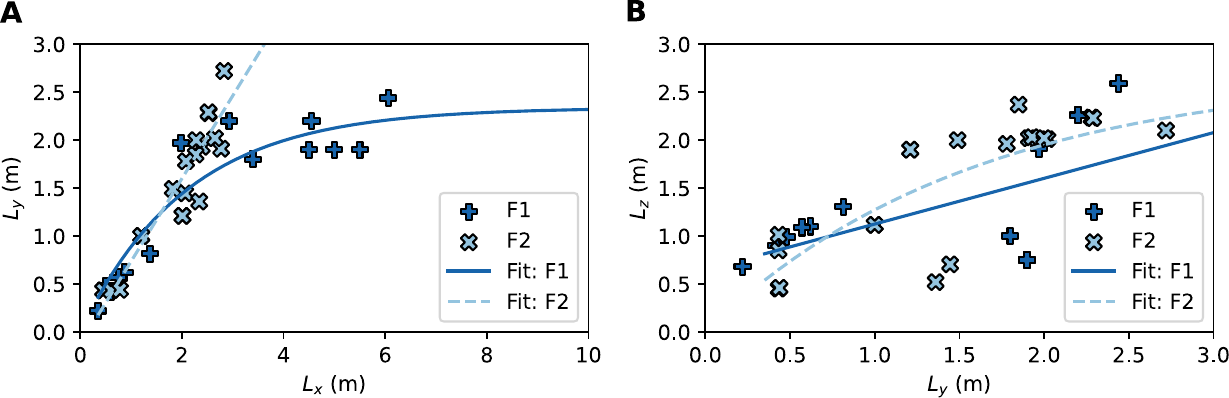}
\caption{Co-dependencies between (A) length and width and (B) length and height in urban furniture (UFD-F1) and private fixtures (UFD-F2).}
\label{Fig:SimpleGeom-furniture_L}
\end{figure}

An analogous analysis is conducted for UFD-F. For UFD-F1 and UFD-F2, likewise UFD-V, co-dependencies are revealed for the bounding box dimensions (shown in Fig. \ref{Fig:SimpleGeom-furniture_L}A,B). Urban furniture (UFD-F1) generally has an upper bound of 6 m for $L_x$, while private fixtures (UFD-F2) remain at about half that size (\mbox{$\approx$ 3.4 m}).

Table \ref{tab:Ly-a-b-RMSE-vehicle_L} provides the error-minimizing functional relationships (Eq. \ref{eq:correlation}) for $L_y$ and $L_z$ for each UFD-F, which are also illustrated in Fig. \ref{Fig:SimpleGeom-furniture_L}A,B.

\begin{figure}[H]
\centering
\includegraphics[width=1\textwidth]{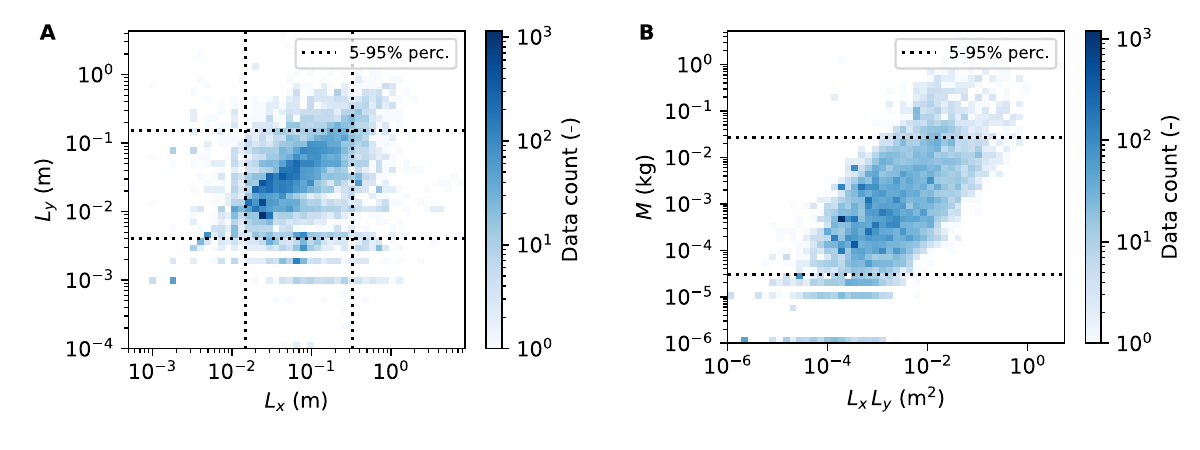}
\caption{Main characteristics derived from the reanalysis of the litter data of \cite{deLange2023litter}. (A) Main length ($L_x$) and width ($L_y$) and (B) mass ($M$) distribution. Linear fit both using a error-minimizing \citep[as implemented in][]{Harris2020array} and robust (Theil-Sen estimator, as implemented by \cite{Virtanen2020scipy}) strategy to depict potential structural relationships between key properties.}
\label{Fig:deLange2023data_L}
\end{figure}

Lastly, UFD-H (encompassing categories DC, DM, DP, DW, and DO in Table \ref{tab:Classification_UFDs}) are considered. 
As commercial catalogues with dimensions and characteristics are not available for UFD-H (mainly: DC, DM), an order of magnitude analysis (Table \ref{tab:debris_param_geom_L}) is performed to establish their dimensions ($L_x$, $L_y$, and $L_z$), albeit with a larger degree of uncertainty than values considered for UFD-V and UFD-F.
To be representative of the debris carried by floods, we have reanalyzed the over 16,000 litter elements inventoried by \cite{deLange2023litter} (Fig. \ref{Fig:deLange2023data_L}A,B).
Based on that data, we suggest triangular probability density functions for $L_x$ and $L_y$ covering of 90\% confidence intervals for min-max and median assumed as the mode of the $\mathcal{T}$ distribution, see Fig. \ref{Fig:deLange2023data_L}A. $L_z$ of DP and DO is chosen as a fraction of $L_y$, therefore assuming that the UFD rests in the most stable configuration.
For wood dimensions, values of Table \ref{tab:debris_param_geom_L} are inspired in the data review of \cite{Gurnell2013wood},

\begin{table}[H]
    \centering
    \begin{tabular}{ p{0.3cm} p{1.5cm} p{3cm} p{3cm} p{3cm} }
        \hline
        ID        & Drifter              & $L_x$ (m)             & $L_y$ (m)             &        $L_z$ (m) \\
        \hline
        DC        & Construction        & $\mathcal{U}$[0.05,1] & $\mathcal{U}$[0.05,1] &        $\mathcal{U}$[0.05,1]\\
        DM        & Metal        & $\mathcal{U}$[0.05,1] & $\mathcal{U}$[0.05,1] &        $\mathcal{U}$[0.05,0.20]\\
        DP        & Plastic        & $\mathcal{T}$[0.015,0.045,0.325] &         $\mathcal{T}$[0.004,0.025,0.154] & $\mathcal{U}$[0.1$L_y$,$L_y$] \\
        DW        & Wood        & $\mathcal{T}$[0.10, 1, 9.0] &         $\mathcal{T}$[0.02,0.15,0.50] & $\mathcal{U}$[0.10,0.50]\\
        DO        & Others         & $\mathcal{T}$[0.015,0.045,0.325] &         $\mathcal{T}$[0.004,0.025,0.154] & $\mathcal{U}$[0.1$L_y$,$L_y$]\\
        \hline
    \end{tabular}
    \caption{Parametrization of the stochastic term ($y'$) of Eq. \ref{eq:correlation} for heterogeneous UFD characteristics; $\Upsilon \equiv$ 0. Plastics (DP) and others (DO) length scales ($L_x$ and $L_y$) based on reanalysis of the data of \cite{deLange2023litter}.}
    \label{tab:debris_param_geom_L}
\end{table}

For all UFD samples, once independent random values of $L_x$, $L_y$, and $L_z$ --within the predefined ranges-- are generated, these are arranged such that $L_x > L_y > L_z$ to ensure the most stable configuration is considered in our stability analysis.

\subsection{Clearance height ($z_c$)}
\label{sec:clearance}

The clearance height $z_c$ represents a critical geometric parameter as it sets the inundation height threshold where the submerged volume (Section \ref{sec:fV_formula}) —and consequently flotation due to internal voids— 
of an UFD, significantly escalates.

\begin{table}[H]
    \centering
    \begin{tabular}{ p{0.5cm} p{0.5cm} p{1.25cm} p{1.25cm} p{1.25cm} p{1.8cm} }
        \hline
        ID &  $y$ & min($y'$) & mode($y'$) & max($y'$) & $y'$ (PDF)\\
        \hline
        V1 & $z_c$  & 0.0   &   0.0   &   0.0   &   $\mathcal{C}$\\
        V2 & $z_c$  & 0.125   &   {-}   &   0.320   &  $\mathcal{U}$\\
        V3 & $z_c$  & 0.118   &   {-}   &   0.42   &  $\mathcal{U}$\\
        V4 & $z_c$  & 0.257   &   {-}   &   0.323   &   $\mathcal{U}$\\
        V5 & $z_c$  & 0.320*   &   {-}   &   0.740*   &   $\mathcal{U}$\\ 
        \hline
    \end{tabular}
    \caption{Stochastic term for the clearance ($z_c$) values of UFD-V ($\Upsilon \equiv$  0 ). Probabilistic distributions based on observations presented in Figure \ref{Fig:SimpleGeom-vehicles-zc}A,D.
    (*) Based on low-floor buses (min), conventional buses (max) as proposed by \cite{Vuchic2007urban}.}
    \label{tab:zc_distro-vehicles}
\end{table}

\begin{figure}[H]
\includegraphics[width=0.45\textwidth]{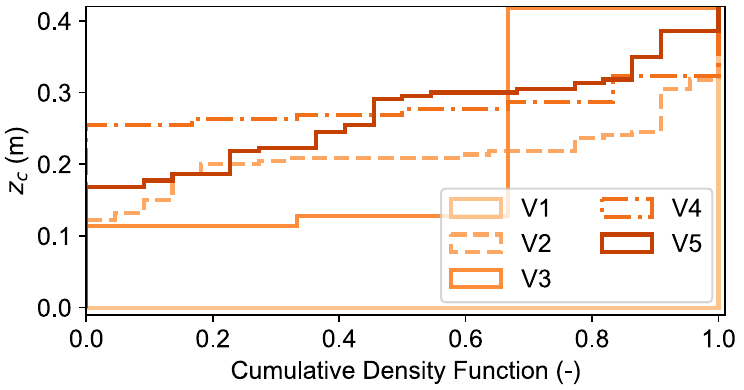}
\caption{Cumulative density function of the clearance height ($z_c$) in vehicles (V1-V5). Note that for UFD-V5, the ground clearance might be visually hidden and we therefore adopt ranges from \cite{Vuchic2007urban}, as clarified in Table \ref{tab:zc_distro-vehicles}.}
\label{Fig:SimpleGeom-vehicles-zc}
\end{figure}

The clearance height of UFD-V ($z_c$) is directly determined from images (obtained from media and social media data available online, featuring side view shots for specific UFD-V models) for a selected set of vehicles (listed in the ESM), and show in Fig. \ref{Fig:SimpleGeom-vehicles-zc}. The $z_c$ value extends from the ground level up to the undercarriage of the UFD-V, with the following two exceptional cases considered:

\begin{enumerate}
    \item UFD-V1 are presumed to be lying down (considering this as their most stable position). Thus, $z_c$ is assumed to be zero. 
    \item For UFD-V5, which include buses and trucks, we note that large voids within the vehicle start at heights greater than the lower chassis levels observed in images \citep[see Fig. 5.3 top of][]{Vuchic2007urban}. In light of the review by \cite{Vuchic2007urban}, we propose a $z_c$ range for UFD-V5 from 0.32 m (low-floor buses) to 0.74 m (conventional buses). From images, we observe that the distribution between the minimum and maximum tends to appear uniform (linear in cumulative terms, see Fig. \ref{Fig:SimpleGeom-vehicles-zc}D), and we therefore extend this PDF to the min-max values extracted from \cite{Vuchic2007urban}.
\end{enumerate}

In the case of UFD-V, the definition of the clearance height is straightforward. However, for urban and private furniture (UFD-F), such objects seldom have wheels, and even when they do, the diameters are just a few centimeters. Thus, for simplicity, we assign a clearance height of $z_c$ = 0 to both UFD-F1 and UFD-F2. For UFD-H (Table \ref{tab:Classification_UFDs}), no wheels are expected, and we embrace the same hypothesis. All the chosen values are presented in Table \ref{tab:zc_distro-vehicles}, with their justifications detailed below.

\subsection{Bulk density ($\rho_b$) or mass ($M$)}
\label{sec:rho_b}
The bulk density of UFDs is a key parameter involved in the calculation of the balance of forces and moments to evaluate the stability of UFDs for the three modes defined in Fig. \ref{Fig:destabilization}.
Density across UFDs, nonetheless, varies largely.
In order to estimate this bulk density, $\rho_b$, we consider Eq. \ref{eq:rho_b}, 
for which the total UFD mass $M$ is necessary, together with the bounding box volume defined by its dimensions of Section \ref{sec:length-scales} ($L_x \, L_y \, L_z$). These variables, readily available in commercial repositories, enable the calculation of $\rho_b$.

\begin{figure}[H]
\includegraphics[width=0.45\textwidth]{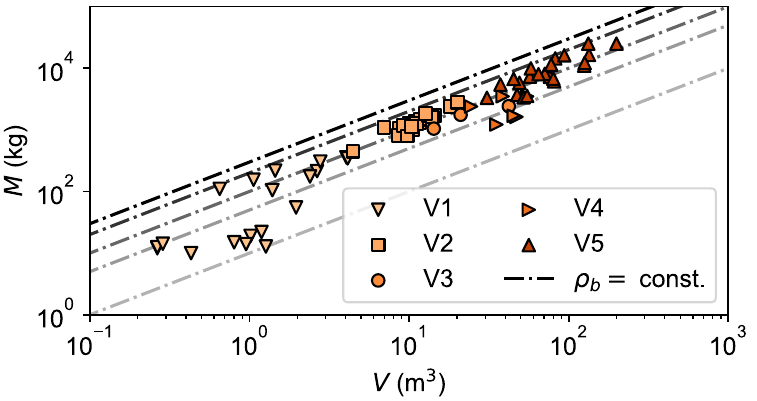}
\caption{Relationship between volume and mass in vehicles (UFD-V1 to UFD-V5).}
\label{Fig:SimpleGeom-vehicles-rho}
\end{figure}

When inspecting the inventory of UFD-V (Figure \ref{Fig:SimpleGeom-vehicles-rho}), a consistent clustering of the bulk density can be observed for different types of vehicles. 
Larger variability is observed in UFD-V1, given the heterogeneity within this category (e.g., this includes bikes and e-scooters as well as grand-tour motorbikes). 
Fig \ref{Fig:SimpleGeom-vehicles-rho} reveals that density is relatively homogeneous across groups UFD-V2 to UFD-V5, except for the UFD-V4, which includes caravans and recreational vehicles  (RVs), where the trend is reversed and bulk density decreases. This fact makes these vehicles relatively more susceptible to flotation. 
In the case of UFD-V4, an increase of volume is associated to an increase on habitability by increasing the internal space; thus leading to a larger increase of volume ($V$) without a significant increase of mass ($M$).

Functional relationships are proposed to predict the bulk density of each UFD-V type (UFD-V2 to UFD-V5, Table \ref{tab:Ly-a-b-RMSE-vehicle_rho}), based on the dimension $L_y$. For UFD-V1, a min-max uniform distribution is assumed covering the range of densities observed. This is based on a decorrelation of the data for this specific UFD. The root mean square error (RMSE) between functional relationships' predictions (Table \ref{tab:Ly-a-b-RMSE-vehicle_rho}) and the data is minimized, and this RMSE is subsequently used to enforce a stochastic component ($y'= \rho_b '$, c.f. Eq. \ref{eq:correlation}) with equivalent standard deviation.

\begin{table}[H]
    \centering
    \begin{tabular}{ p{1cm} p{0.5cm} p{1.0cm} p{1.25cm} p{2.2cm} p{2.2cm} p{2.2cm} p{3cm} }
        \hline
        ID              & $y$ & $x$ & $\Upsilon$ (fit) &  $a$  & $b$ & RMSE & $y'$ (PDF)\\
        \hline
        V1              & $\rho_b$ &  {-} & {-} & {-} & {-} & {-} &  $\mathcal{U}$[10.1, 169.0] kg m$^{-3}$\\
        V2              & $\rho_b$ &  {$L_y$} & Linear & 56.856 kg m$^{-4}$ & 20.198 kg m$^{-3}$ & 14.858 kg m$^{-3}$ &  $\mathcal{N}$\\
        V3              & $\rho_b$ &  {$L_y$} & Linear & 16.686 kg m$^{-3}$ & 38.175 kg m$^{-4}$ & 10.088 kg m$^{-3}$ &  $\mathcal{U}$\\
        V4              & $\rho_b$ &  {$L_y$} & Linear & -109.005 kg m$^{-3}$ & 301.323 kg m$^{-4}$ & 16.462 kg m$^{-3}$ &  $\mathcal{U}$\\
        V5              & $\rho_b$ &  {$L_y$} & Linear & 144.172 kg m$^{-3}$ & -233.533 kg m$^{-4}$ & 32.620 kg m$^{-3}$ &  $\mathcal{U}$\\
        \hline
        F1  & $M$ & {$L_x \cdot L_y$} &  Saturation    & 2334.068 kg & 0.0700 m$^{-2}$ & 197.460 kg &  $\mathcal{U}$\\
        F2  & $M$ & {$L_x \cdot L_y$} &  Linear    & 15.386 kg/m$^2$ & 30.453 kg & 50.242 kg &  $\mathcal{N}$\\
        \hline
    \end{tabular}
    \caption{Structural ($\Upsilon$) and stochastic ($y'$) relationships (Eq. \ref{eq:correlation}) for the bulk density ($\rho_b$) or mass ($M$), based on data analysis presented in Figure \ref{Fig:SimpleGeom-vehicles-rho}A,B, Figure \ref{Fig:SimpleGeom-furniture_rho}. 
    }
    \label{tab:Ly-a-b-RMSE-vehicle_rho}
\end{table}

\begin{figure}[H]
\includegraphics[width=0.45\textwidth]{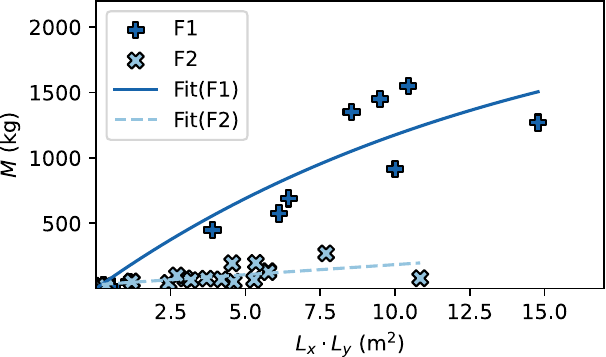}
\caption{Structural relationship between UFD-F mass and the bounding box base area, including urban furniture (UFD-F1) and private fixtures (UFD-F2).}
\label{Fig:SimpleGeom-furniture_rho}
\end{figure}

The mass of UFD-F, however, is found to be better explained by the bounding box horizontal area, defined by the product $L_x \cdot L_y$ (Fig. \ref{Fig:SimpleGeom-furniture_rho}).
This correlation may be attributed to the stability considerations, as a larger footprint (bounding box base area) often corresponds with greater height (Fig. \ref{Fig:SimpleGeom-furniture_rho}B).
Within the furniture category, it is also clear that urban furniture (UFD-F1) is considerably denser than private fixtures (UFD-F2), as shown in Fig. \ref{Fig:SimpleGeom-furniture_rho}.

For heterogeneous UFDs, we combine the raw material density ($\rho_s$) with a plausible volume fraction to define the bulk density ($\rho_b = \rho_s f_V$ ), which then allows obtaining the mass ($M$), as defined by Eq. \ref{eq:rho_b}. 
Plastics (DP) and others (DO) are an exception, since they can hold a very wide  range of densities \citep{vanEmmerik2020}, as well as a wide range of internal voids. 
To accurately represent the debris found in waterways, we have reanalyzed the over 16,000 litter elements inventoried by \cite{deLange2023litter} (Fig. \ref{Fig:deLange2023data_L}B). Based on statistical properties of that data, we define the parameters of a $\mathcal{T}$ distribution as follows: min-max based on 5 to 95\% percentiles and mode assumed as the median of the data (Fig. \ref{Fig:deLange2023data_L}B).
This is also extended to DO, under the consideration that both groups represent litter.
For wood, table \ref{tab:debris_param_dynamics_rho} presents the parametrization of the PDFs describing the raw density ($\rho_s$) of each material and the $f_V$ (Table \ref{tab:debris_param_geom_fV}), later discussed in Section \ref{sec:fV_formula}, which allow the estimation of the bulk density via $\rho_b = \rho_s f_V$.\\

\begin{table}[H]
    \centering
    \begin{tabular}{ p{0.3cm} p{1.5cm} p{3cm} p{1.5cm} p{2.5cm} }
        \hline
        ID        & Drifter              & $\rho_s$ (kg/m$^3$)   &        $\rho_b$  (kg/m$^3$)  & $M$ (kg) \\
        \hline
DC & Construction & $\mathcal{T}$[1700,2400,2500] & $= \rho_s \cdot f_V$ & $= \rho_b \cdot V$ \\
DM & Metal & $\mathcal{U}$[7500,8000] & $= \rho_s \cdot f_V$ & $= \rho_b \cdot V$ \\
DP & Plastic & {} & {} & $\mathcal{T}$[0.03,0.55,26.5]$\cdot$10$^{-3}$ \\
DW & Wood & $\mathcal{N}$[713.39, 207.27] & $= \rho_s \cdot f_V$ & $= \rho_b \cdot V$ \\
DO & Others & {} & {} & $\mathcal{T}$[0.03,0.55,26.5]$\cdot$10$^{-3}$ \\
        \hline
    \end{tabular}
    \caption{Parametrization of heterogeneous UFD massive characteristics. Plastics (DP) and others (DO) mass ($M$) based on reanalysis of the data of \cite{deLange2023litter}, as presented in Fig. \ref{Fig:deLange2023data_L}B. Wood based on the reanalysis of the data of \cite{chave2009towards}, as presented in Fig. \ref{Fig:wood_distr}.}
    \label{tab:debris_param_dynamics_rho}
\end{table}

\begin{figure}[H]
\includegraphics[width=0.45\textwidth]{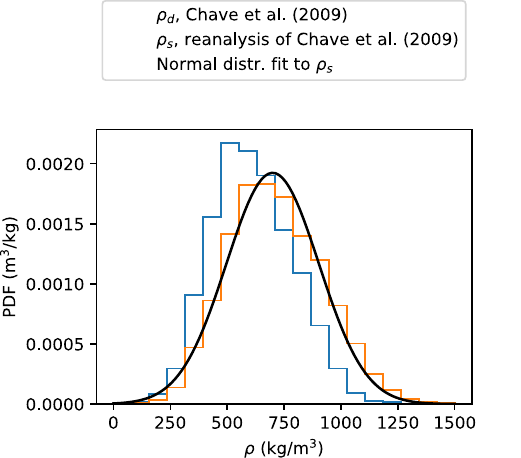}
\caption{Reanalysis of the data of \cite{chave2009towards} and calculation of wet density through Eqs. \ref{Eq:rhos_rhod} and \ref{Eq:WC}, assuming $\mathcal{T}$[0.04, 0.15, 0.30]. 
Oven-dry wood density ($\rho_d$) of 16,467 tree species measurements across 1,683 genera and 191 families based on the data of \cite{chave2009towards}.}
\label{Fig:wood_distr}
\end{figure}

Defining wood density is a complex task as it relies on various factors, including tree species, part, age, and season. \citet{chave2009towards} provides a comprehensive dataset of 16,467 density measurements across 1,683 genera and 191 families. This dataset suggest a symmetric distribution with a mean of 613.4 kg/m$^3$ and a standard deviation of 175.9 kg/m$^3$ (mode 560 kg/m$^3$, maximum 1390 kg/m$^3$, and minimum 80 kg/m$^3$) for the wood density, as depicted in Figure \ref{Fig:wood_distr}.

The density referred to here is the oven-dry density ($\rho_d$), which represents the mass after water has been evaporated divided by the original (green) volume. The actual (green) density ($\rho_s$) can be calculated as follows:

\begin{equation}
\rho_s = \rho_d \cdot (1 + WC)
\label{Eq:rhos_rhod}
\end{equation}

where $WC$ is the water content, defined as a function of green mass ($M$) and oven-dry mass ($M_d$) as follows:

\begin{equation}
WC = \frac{M - M_d}{M_d}
\label{Eq:WC}
\end{equation}

We take the dry density of \cite{chave2009towards} and assume that the WC varies between 0.04 and 0.30, with a most common value (mode) at around 0.15 (WC $\sim \mathcal{T}$[0.04, 0.15, 0.30]). The result is presented in Fig. \ref{Fig:wood_distr} together with a Gaussian fit for $\rho_d$ that is later used to estimate wood density in our probabilistic analysis. This wood density estimation (mean: 713.39 kg, std: 207.27 kg) lays in between densities estimated by \cite{ruiz2016wood} for green wood (mean: 800 kg, std: 170 kg) and instream wood (mean: 660 kg, std: 200 kg).



\subsection{Fraction of effective drag area ($f_A$)}
\label{sec:fa}

The effective area of an UFD that is exposed to the flow depends on the water depth $h$. With increasing water level, larger areas will become wet on the impinging face (upstream UFD face, $A'$ in Fig. \ref{Fig:BoundingBox_fA_fV}D). For convenience, we describe the frontal area of the UFD through the parameter $f_A$ (Eq. \ref{eq:f_A}), which relates the effective drag area $A'$ to the flow-exposed area of the bounding box; i.e., $L_x h$.

The estimation of $f_A$ for different UFDs follows different strategies depending on their category. 
For typified UFDs, UFD-V and UFD-F, we implement the following image-based analysis procedure over selected items of the UFD-inventory (see ESM for completeness):

\begin{enumerate}
    \item We collect images of vehicles and furniture from online commercial repositories. These images are selected only for side looking angles, corresponding to the $x-z$ plane (flow-exposed area of the bounding box).
    \item We binarize the images, removing any potential background (commonly white) and enhancing the UFD (bringing it to black). This is checked visually for adequacy.
    \item We assume different elevation levels $h$ at which the water level may reach (c.f. Fig. \ref{Fig:BoundingBox_fA_fV}C,D). For a given $h$, we count the number of black pixels underneath, and divide over the total number of pixels (black and white together). This ratio provides a direct estimation of $f_A$.
\end{enumerate}

This procedure facilitates the empirical determination of $f_A$ for various UFD-V and UFD-F types across an extensive range of potential water levels to which the UFD might be exposed. From the obtained data, a polynomial expression is fitted to facilitate the evaluation of $f_A$ at any given depth during the application of the model of Section \ref{sec:MechModel}. This polynomial fit is obtained for each group within UFD-V and UFD-F, as defined in Table \ref{tab:Classification_UFDs}, and follows the form:

\begin{equation}
    f_A (h) \approx \sum_{i=0} ^3 a_i \left( h/L_z \right) ^i 
    \label{eq:fA_poly}
\end{equation}
where $a_i$ are error-minimizing coefficients (presented in Table \ref{tab:fA_coeffs}).\\

\begin{table}[H]
    \centering
    \begin{tabular}{ p{0.3cm} p{1cm} p{1cm} p{1cm} p{1cm} p{1cm}}
        \hline
        ID  & $a_0$ & $a_1$ & $a_2$ & $a_3$ & RMSE \\
        \hline
        V1  & 0.186 & 2.094 & -3.001 & 1.274 & 0.192 \\
        V2  & -0.010& 2.761 & -2.905& 0.846 & 0.039 \\
        V3  & -0.017& 2.926 & -3.390& 1.254 & 0.070 \\
        V4  & -0.036& 3.179 & -4.037& 1.694 & 0.051 \\
        V5  & -0.008& 2.543 & -2.546& 0.840 & 0.083 \\
        \hline
        F1  & 0.431 & 1.558 & -2.232& 1.087 & 0.183 \\
        F2  & 0.530 & 2.441 & -4.135& 2.122 & 0.121 \\
        \hline
    \end{tabular}
    \caption{Coefficients for the $f_A$ fit (Eq. \ref{eq:fA_poly}) obtained from image processing of pictures of side UFDs-V and UFDs-F with plain background.}
    \label{tab:fA_coeffs}
\end{table}

\begin{figure}[H]
\includegraphics[width=0.475\textwidth]{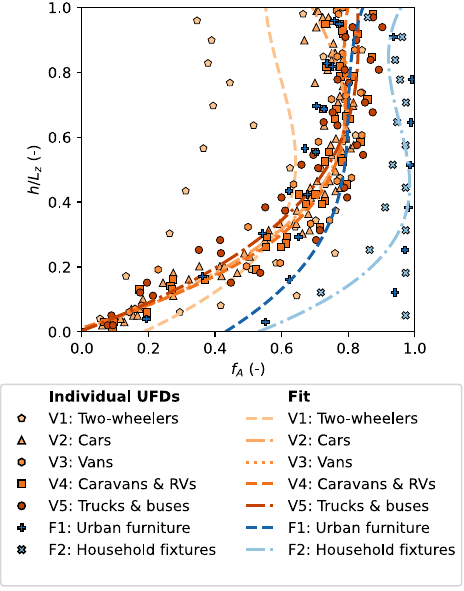}
\caption{Fraction of effective drag area ($f_A$) and polynomial fit (Eq. \ref{eq:fA_poly}) per UFD-V and UFD-F subcategories, based on polynomial coefficients presented in Table \ref{tab:fA_coeffs}.}
\label{Fig:fa}
\end{figure}

The resulting $f_A$ curves are presented in Fig. \ref{Fig:fa}, together with the data retrieved from the image-based procedure proposed.

We also estimate the RMSE of Eq. \ref{eq:fA_poly} for UFDs of each type of vehicle (Table \ref{tab:fA_coeffs}) which, in our probabilistic framework, allows to consider deviations from the expected trend (given by Eq. \ref{eq:fA_poly}) through $f_A'$ in the following relation:

\begin{equation}
    f_A (h) =  \min( (1 + f_A') \cdot \sum_{i=0} ^{3} a_i \left( h/L_z \right) ^i , 1)
    \label{Eq:f_A_implementation}
\end{equation}
 
The RMSE of the fitting of Eq. \ref{eq:fA_poly} to the empirical data is used to model $f_A'$, which adopts a uniform PDF that holds the same standard deviation as the RMSE; i.e., $f_A'\sim \mathcal{U}$ [$-\sqrt{3}$ RMSE, $\sqrt{3}$ RMSE], with null-mean value. 

In our probabilistic framework (Section \ref{sec:MC_analysis}), a $f_A'$ value is generated for each simulation and kept constant across the whole water column; 
i.e., each UFD generated in the Monte Carlo analysis  has a constant deviation from the mean trend defined by Eq. \ref{Eq:f_A_implementation} and the $a_i$ values that are characteristic of each group within UFD-V and UFD-F (Table \ref{tab:fA_coeffs}).  
For instance, if a random $f_A'$ is generated with a value of 0.05 for one stability calculation, all $f_A$ will be 5\% larger than the empirically-calibrated Eq. \ref{eq:fA_poly} of that type of vehicle. 
If, due to randomness of $f_A'$, $f_A (h)$ of Eq. \ref{Eq:f_A_implementation} takes a value above one, it is bounded down to unity.

As shown in Figure \ref{Fig:fa}, excluding UFD-V1 (bikes, motorbikes, and e-scooters), the $f_A$ curves for various UFD-V types remain remarkably consistent. 


For UFD-H, $f_A$ values are assigned in a broad range to reflect considerable uncertainties. These values often start around 0.4 to 0.6, can take values considering specific shapes (circular areas, 0.78), and may go up to nearly filling the entire flow-exposed area of the bounding box ($f_A$ of approximately 0.95). This higher $f_A$ value is generally expected when the UFD-H shapes are more stylized, such as bars or bricks. The proposed $f_A$ values for each UFD-H class are presented in Table \ref{tab:debris_param_geom_fA}.

\begin{table}[H]
    \centering
    \begin{tabular}{ p{0.3cm} p{2cm} p{2cm}  }
        \hline
        ID        & Drifter              & $f_A$ (-) \\
        \hline
        DC        & Construction        & $\mathcal{U}$[0.6,0.95] \\
        DM        & Metal               & $\mathcal{U}$[0.8,0.95]\\
        DP        & Plastic             & $\mathcal{U}$[0.4,0.6]\\
        DW        & Wood                & $\mathcal{U}$[0.8,0.95]\\
        DO        & Others              & $\mathcal{U}$[0.4,0.6]\\
        \hline
    \end{tabular}
    \caption{Parametrization of heterogeneous UFD characteristics (I). Plastics (DP) and others (DO) frontal area fraction ($f_A$).}
    \label{tab:debris_param_geom_fA}
\end{table}

\subsection{Fraction of submerged volume ($f_V$)}
\label{sec:fV_formula}
The buoyancy force is proportional to the submerged volume (refer to Eq. \ref{eq:S_z}). In our model, we represent internal volumes using the factor $f_V$ (as defined in Eq. \ref{eq:f_V}). This factor represents the fraction of the bounding box volume submerged ($L_x L_y h$) contributing to buoyancy.

For $f_V$, we introduce a two-step function to reflect changes in UFD-V's buoyancy with increasing depth:

\begin{equation}
    f_V \approx
    \begin{cases}
      f_{V,\text{1}} & \text{if } h \leq z_c\\
        \left( f_{V,\text{1}} \cdot z_c + f_{V,\text{2}} \cdot [h - z_c]
        \right) / h & \text{if } h > z_c\\
    \end{cases}      
    \label{Eq:f_V_function}
\end{equation}

In this function, $f_{V,\text{1}}$ accounts for the buoyancy contribution below the clearance height $z_c$, which we assume to be small and constant for $h < z_c$. The term $f_{V,\text{2}}$ represents the sudden buoyancy increase when water level surpasses the under-chassis.

We select $f_{V,\text{1}}$ to represent the volume occupied by wheels under the clearance ($f_{V,\text{1}}\sim$5\%, as shown in Table \ref{tab:UFDs_fV}). For $f_{V,\text{2}}$ of cars (UFD-V2), we decide its value based on calibration of the mechanistic model on real-scale data (described in section \ref{sec:calibration}). For other vehicles, we rely on simple geometrical assessments based on images, which may be regarded as approximate. This attribute carries higher uncertainty compared to other physical quantities, warranting a uniform PDF adoption, which endorses large uncertainty across a give min-max range. The adopted values for $f_{V,\text{2}}$ are introduced and briefly justified in Table \ref{tab:UFDs_fV}.

\begin{table}[H]
    \centering
    \begin{tabular}{ p{2.5cm} p{1.5cm} p{1.5cm} p{9.5cm} }
        \hline
        UFD & $f_{V1}$ &  $f_{V2}$ & Comments\\
        \hline
        V1    & -  &   $\mathcal{U}$[1.2$\psi$,3$\psi$]   &   *Assuming that all volume is due to aluminium (2,700 kg/m$^3$), the mass $M$ is converted to a volume ($V_{\text{solid}}$), then: $\psi = V_{\text{solid}}/V$. We then assume an extra 20\% volume due to light components (plastic) and larger voids, which we assume that can go up to 3$\psi$.\\
        V2   & 0.05   & $\mathcal{U}$[0.30,0.50]   &   Based on the reanalysis of the data of \cite{Smith2019full}, \cite{Kramer2016safety} and \cite{Xia2014criterion}.\\
        V3, V4, V5   & 0.05   &  $\mathcal{U}$[0.60,0.80] &   Large closed volumes, assumed water-tight.\\
        \hline
    \end{tabular}
    \caption{Values of the void fraction coefficient $f_{V1}$ and $f_{V2}$ (Eq. \ref{Eq:f_V_function}).}
    \label{tab:UFDs_fV}
\end{table}

In the case of UFD-F, we derive the value of $f_{V,2}$ (as there's no clearance, hence no $f_{V,1}$) by linking it to $f_A$. We make the approximation $f_V \approx f_A^{3/2}$ (prismatic shaped).
For UFD-H, we assume sensible values that span a wide range of plausible possibilities. This incorporates larger uncertainty but ensures that we cover realistic combinations. The following assumptions were incorporated when defining UFD-H distributions of $f_V$: for wood, we typically assume cylindrical shapes (\mbox{$f_V \approx$ 0.78}). However, wood can sometimes take larger volume fractions without fully becoming prismatic (upper limit of \mbox{$f_V$ < 0.95}). Since wood can also break into anisotropic pieces, we consider a lower limit of \mbox{$f_V$ > 0.50}. These considerations justify a triangular distribution for $f_{V}$ for wood ($\mathcal{T}$[0.5,0.78,0.95]), as presented in Table \ref{tab:debris_param_geom_fV}.\\

\begin{table}[H]
    \centering
    \begin{tabular}{ p{0.3cm} p{2cm} p{2.3cm}  }
        \hline
        ID        & Drifter             & $f_{V}$ (-) \\
        \hline
        DC        & Construction        & $\mathcal{T}$[0.5,0.78,0.95] \\
        DM        & Metal               & $\mathcal{T}$[0.5,0.78,0.95] \\
        DP        & Plastic             & $\mathcal{T}$[0.5,0.78,0.95] \\
        DW        & Wood                & $\mathcal{T}$[0.5,0.78,0.95] \\
        DO        & Others              & $\mathcal{U}$[0.4,0.6] \\
        \hline
    \end{tabular}
    \caption{Submerged volume fraction parametrization of UFD-H.}
    \label{tab:debris_param_geom_fV}
\end{table}

\subsection{Friction coefficient ($\mu$)}
\label{sec:mu}

The friction coefficient describes the relationship between normal and frictional forces in the UFD-floor contact (see Eq. \ref{eq:R_y}).
This coefficient is varies according to the material and roughness properties of the two contact elements.
Based on literature review, we define a minimum, mode, and maximum value for each material pair (characteristic of different UFD-floor types), always factoring in wet contact conditions. We list these values in Table \ref{tab:friction_coefficients}.\\

\begin{table}[H]
    \centering
    \begin{tabular}{ p{2cm} p{3cm} p{0.75cm} p{0.75cm} p{0.75cm} p{1.25cm} p{5.0cm} }
        \hline
        Material 1 & Material 2 & min($\mu$) & mode($\mu$) & max($\mu$) & UFD & Reference(s)\\
        \hline
        Tire &   Asphalt/concrete   & 0.30   &   0.50   &   0.75   & V1--V5, F1, F2, DP, DO &  Min: \cite{Smith2019full}, mode: \cite{Wong2008book}, max: \cite{Smith2019full}.\\
        Rock/concrete &   Asphalt/concrete   & 0.55   &   0.87   &   1.10   & F1, F2, DC &  Min: \cite{Zhao2020friction}, mode: \cite{Zhao2020friction}, max: \cite{Zhao2020friction}\\
        Wood &   Asphalt/concrete   & 0.60   &   0.80   &   0.90   &  F1, F2, DW & Min: \cite{Jaaranen2020frictional}, mode: \cite{Gorst2003friction}, max: \cite{Gorst2003friction}\\
        Metal &   Asphalt/concrete   & 0.62   &   0.65   &   0.67   &  DM & Min: \cite{Rabbat1985friction}, mode: \cite{Rabbat1985friction}, max: \cite{Rabbat1985friction}\\
        \hline
    \end{tabular}
    \caption{Friction coefficients for different pairs of materials under wet contact conditions. Selection of UFDs potentially lying in each material-pairs configuration.}
    \label{tab:friction_coefficients}
\end{table}

To model the friction coefficient ($\mu$), we make the assumption that all UFD-V types are placed over asphalt/concrete (see Table \ref{tab:friction_coefficients}). We represent this via a triangular distribution $\mathcal{T}$[0.30, 0.50, 0.75].\\

For UFD-F, we consider a reasonably wide range within the material pairs, using a uniform distribution: $\mathcal{U}$[0.30, 0.90].
For UFD-H, we adopt a triangular distribution, grounded on defined ranges dependent on the pair of materials in contact (refer to Table \ref{tab:friction_coefficients}).
Consequently, the distributions adopted for UFD-H are described in Table \ref{tab:debris_param_dynamics_mu}.\\

\begin{table}[H]
\centering
\begin{tabular}{ p{0.3cm} p{1.8cm} p{2.5cm} }
\hline
ID        & Drifter           &  $\mu$ (-) \\
\hline
DC  &  Construction  &   $\mathcal{T}$[0.55,0.87,1.10]  \\
DM  &  Metal  &  $\mathcal{T}$[0.62,0.65,0.67]  \\
DP  &  Plastic  & $\mathcal{T}$[0.62,0.65,0.67]  \\
DW  &  Wood  &   $\mathcal{T}$[0.62,0.65,0.67]  \\
DO  &  Others  &  $\mathcal{T}$[0.62,0.65,0.67]  \\
\hline
\end{tabular}
\caption{Parametrization of friction factors for UFD-H.}
\label{tab:debris_param_dynamics_mu}
\end{table}

\subsection{Drag coefficient ($C_D$)}
\label{sec:CD_param}
The drag coefficient directly scales the drag force exerted by the water upon the vehicle (Eq. \ref{eq:S_y}).
Some drag coefficients are known for vehicles under flooding conditions \citep[see for instance][]{Smith2019full}.
In this investigation, the drag coefficients of UFD-V, UFD-F and UFD-H are assimilated from simpler geometrical shapes \citep[see Table 7.2-7.3 of ][]{White2016}.
A broad review of $C_D$ for arbitrary shapes is presented in Table \ref{tab:UFDs_CD}, together with the justification on their assimilation to specific types of UFD.

\begin{table}[H]
    \centering
    \begin{tabular}{ p{2.8cm} p{1.2cm} p{1.2cm} p{1.2cm} p{10cm} }
        \hline
        UFD & min($C_D$) & mode($C_D$) & max($C_D$) & Comments\\
        \hline
        V1: Bikes/e-scooters   & 0.57   &   0.82   &   1.07   &  Based on the drag of a cylinder with length/diameter ratio of ten.\\
        V1: Motorbikes   & 0.82   &   1.17   &   1.52   &  Mode value based on the drag of a disk, max-min uncertainty (\%) assimilated from cars.\\
        V2: Cars   & 0.98   &   1.40   &   1.83   &   Min: \cite{Smith2019full}, mode: \cite{Wong2008book}, max: \cite{Smith2019full}. Variability is also representative of other studies on cars and simplified blunt shapes.\\
        V3, V4, V5: Vans, caravans, trucks   & 1.08   &   1.54   &   2.01   &   Assimilated from cars and increased by 10\% since shapes are less hydrodynamic.\\
        \hline
        F1: Bins   & 0.64   &   0.68   &   0.72   &  Cylindrical shape, with length to diameter ratios of one, two and three.\\
        F1: Dumpsters   & 1.07   &   1.18   &   1.2   &   Based on a cube (min), a plate with aspect ratio one (mode) and two (max).\\
        F2: Water/gas tanks   & 0.64   &   0.68   &   0.72   &  Cylindrical shape, with length to diameter ratios of one, two and three.\\
        F2: Garden shed   & 1.07   &   1.18   &   1.2   &   Based on a cube (min), a plate with aspect ratio one (mode) and two (max).\\
        \hline
        Rigid plastics   & 1.07   &   1.18   &   1.2   &   Based on a cube (min), a plate with aspect ratio one (mode) and two (max).\\
        Deformable plastics    & 1.07   &   1.18   &   1.20   &   Based on a cube (min), a plate with aspect ratio one (mode) and two (max).\\
        Metal structures    & 1.07   &   1.30   &   1.50   &   Based on a cube (min), a plate with aspect ratio ten (mode) and twenty (max).\\
        Natural wood    & 0.50   &   0.82   &   1.20   &   Min and max based on pine and spruce \citep{Johnson1982trees} and mode based on cylinder with length to diameter ratio of ten.\\
        Construction (wood)   & 1.07   &   1.30   &   1.50   &   Based on a cube (min), a plate with aspect ratio ten (mode) and twenty (max).\\
        Construction (others)    & 1.07   &   1.18   &   1.20   &   Based on a cube (min), a plate with aspect ratio one (mode) and two (max).\\
        \hline
    \end{tabular}
    \caption{Drag coefficients proposed for different UFDs (vehicles, containers and litter). Coefficients based on simplified geometrical shapes are extracted from \cite{White2016}. Flow impinging traverse to the vehicle. Bikes, motorbikes and scooters are considered laying on the floor.}
    \label{tab:UFDs_CD}
\end{table}

In our analysis, a certain additional variability is allowed for $C_D$ to account for a wide range of uncertainties beyond the shape assimilation.
A triangular probability distribution is assumed for $C_D$ in all UFD-V and UFD-F, considering a minimum, a mode and a maximum value (Table \ref{tab:finalCDs}).
The same rationale is used to inform the selection of the probabilistic distributions for UFD-H.
However, for plastic (UFD-DP) and other debris (UFD-DO), a uniform PDF is chosen thereby acknowledging a larger uncertainty in the shapes defining the drag.
When a UFD could be assimilated to several geometric forms, we take the widest range comprehended by these several shapes' values of $C_D$, hence transferring larger uncertainty to the resulting stability curves' predictions. Drag coefficients for UFD-H are presented in Table \ref{tab:debris_param_dynamics_CD}.\\

\begin{table}[H]
    \centering
    \begin{tabular}{ p{1.6cm} p{1.2cm} p{1.2cm} p{1.2cm} p{1.8cm} }
        \hline
        ID &  min($C_D$) & mode($C_D$) & max($C_D$) & Probabilistic distribution\\
        \hline
        V1 & 0.58   &   1.00   &   1.52   &  $\mathcal{T}$\\
        V2 & 0.75*   &   1.40   &   1.70*  &  $\mathcal{T}$\\
        V3, V4, V5 & 1.08   &   1.54   &   2.01   &  $\mathcal{T}$\\
        \hline
        F1, F2 & 0.64   &   0.92   &   1.2   &   $\mathcal{T}$\\
        \hline
    \end{tabular}
    \caption{Drag coefficients ($C_D$) based on real data and simplified geometrical shapes assimilation, see Table \ref{tab:UFDs_CD}.
    }
    \label{tab:finalCDs}
\end{table}

\begin{table}[H]
\centering
\begin{tabular}{ p{0.3cm} p{1.8cm} p{2.5cm} }
\hline
ID        & Drifter           &  $C_D$ (-) \\
\hline
DC  &  Construction  &  $\mathcal{T}$[1.07,1.18,5.0] \\
DM  &  Metal  &  $\mathcal{T}$[1.07,1.3,1.5]  \\
DP  &  Plastic  &  $\mathcal{U}$[1.0,5.0]  \\
DW  &  Wood  &  $\mathcal{T}$[0.5,1.0,1.5] \\
DO  &  Others  &  $\mathcal{U}$[1.0,5.0] \\
\hline
\end{tabular}
\caption{Parametrization of heterogeneous UFD characteristics (II). Plastics (DP) and others (DO) mass ($M$) based on reanalysis of the data of \cite{deLange2023litter}.}
\label{tab:debris_param_dynamics_CD}
\end{table}

\noappendix       

\authorcontribution{Following CRediT (Contributor Roles Taxonomy), the following authors have contributed to the following tasks. Conceptualization: DV, AB, MF; Data curation: AB; Formal Analysis: DV;  Investigation: DV, AB; Methodology: DV, AB, MF; 
Resources: MF; Software: DV, AB; Validation: DV, AB; Visualization: DV, AB; Writing -- original draft: DV, AB, MF; writing -- review \&  editing: DV, AB, MF} 

\competinginterests{The authors declare that they have no conflict of interest.} 

\codedataavailability{The codes producing the stability curves (as well as all the stability curves) are available under: \url{https://github.com/davahue/UFD-stability}. The inventory of UFDs informing the parameters' characterization is also included in that repository.}



\begin{acknowledgements}
The authors are grateful for the discussions they had with Dr. Isabella Schalko regarding wood properties and mobilization.
This research was partially supported by the Erasmus+ STA program (EU), the Program to Support Research \& Development of Universitat Politècnica de València (PAID--06--22), its UPV2030 Strategic Plan's Academic Career Support Program for Professors (PACAP--2022), and the Program for Emerging Research Groups of the Government of Valencia, Spain (CIGE/2022/7).
Figure 1.A was reproduced with AFP' permission.
Figure 1.B was reproduced with AP's permission.
Figure 1.C was reproduced with Getty Images' permission.
\end{acknowledgements}







\bibliographystyle{copernicus}
\bibliography{main.bib}

\end{document}